\documentclass{aastex63}
\usepackage{amsmath, amssymb}

\newcommand{\Pp}{$P^\prime$\ }

\received{January 29, 2021}
\revised{March 8, 2021; April 26, 2021; October 4, 2021}
\accepted{December 3, 2021}
\submitjournal{The Astrophysical Journal}

\shorttitle{Brown Dwarf and YSO Disk Polarization}
\shortauthors{Clemens et al.}

\begin{document}

\title{Near-Infrared Polarization from Unresolved Disks Around \\
Brown Dwarfs and Young Stellar Objects}

\correspondingauthor{Dan Clemens}
\email{clemens@bu.edu}

\author[0000-0002-9947-4956]{Dan P. Clemens}
\affiliation{Institute for Astrophysical Research, Boston University,
    725 Commonwealth Ave, Boston, MA 02215}
\author[0000-0003-2133-4862]{Thushara G.S. Pillai}
\affiliation{Institute for Astrophysical Research, Boston University,
    725 Commonwealth Ave, Boston, MA 02215}
\author[0000-0002-3091-8061]{Anneliese M. Rilinger}
\affiliation{Institute for Astrophysical Research, Boston University,
    725 Commonwealth Ave, Boston, MA 02215}
\author[0000-0001-9227-5949]{Catherine C. Espaillat}
\affiliation{Institute for Astrophysical Research, Boston University,
    725 Commonwealth Ave, Boston, MA 02215}
\begin{abstract}
Wide-field near-infrared (NIR) polarimetry was used to examine disk systems
around two brown dwarfs (BD) and two young stellar objects (YSO) embedded in the Heiles Cloud 2 (HCl2) 
dark molecular cloud in Taurus as well as numerous stars located behind HCl2.
Inclined disks exhibit intrinsic NIR polarization due to scattering
of photospheric light which is detectable even for unresolved systems. 
After removing polarization contributions from magnetically aligned 
dust in HCl2 determined from the background star information, 
significant intrinsic polarization was detected from the disk systems of of one BD (ITG~17) 
and both YSOs (ITG~15, ITG~25), but not from the other BD (2M0444). 
The ITG~17 BD shows
good agreement of the disk orientation inferred from the NIR and from published ALMA dust continuum imaging.
ITG~17 was also found to reside in a 5,200~au wide binary (or hierarchical quad star system) with 
the ITG~15 YSO disk system.
The inferred disk orientations from the NIR for ITG~15 and ITG~17 are 
parallel to each other and perpendicular to the local magnetic field direction.
The multiplicity of the system and the large BD disk nature could have resulted from formation
in an environment characterized by misalignment of the magnetic field and the protostellar disks.
\end{abstract}

\section{Introduction} \label{sec:intro}

Disks around low-mass objects offer unique insights 
into star and planet formation processes \citep[see reviews by][]{Luhman12,Andrews20}, 
especially regarding low-mass star and brown dwarf formation. 
However, such objects are intrinsically faint and their disks less massive, posing challenges for studies of the
dust emission from those disks using radio interferometers.

In addition to emitting thermal radiation, dust in disks also scatters and polarizes central object (and disk) light.
Linear polarimetry observations and modeling of resolved disks \citep[e.g.,][]{Silber00, Apai04, Follette13,Esposito18,Esposito20}  
have been used to constrain disk properties and to identify departures in the polarization patterns, 
and the structures seen in polarized light, as 
bona fide disk features \citep{Avenhaus14,Benisty15,Asensio-Torres16,Garufi18}, perhaps due to planet 
or protoplanet formation. 

Near-infrared (NIR) polarimetry may offer an efficient way to survey, detect, and identify low-mass
star or BD disks for follow-up examination by millimeter (mm) and submillimeter (submm) wavelength interferometers. This current study 
sought to use NIR polarimetry to probe two brown dwarf (BD) disks and two young stellar object (YSO) disks in Taurus to assess the efficacy of this method
and to compare NIR and mm disk orientation findings.

Stars (or BDs) without disks are sources of unpolarized light. Polarization signals detected in the light from distant 
examples of such objects may be evaluated,
for example, to trace magnetic field orientations in intervening dusty material along the line of sight
\citep{Hiltner49a,Hiltner49b,Hall49,Davis51}.
However, those lines of sight may be long and feature complex dust and gas distributions, which complicates
localizing magnetic field properties to particular atomic or molecular clouds of interest.
Foreground and background polarization contributions from dust in the diffuse and 
dense interstellar medium
(ISM) components can become mixed with any target object polarization, masking the
desired signal.
In this study, these polarization contributions were quantified and 
removed in order to isolate the intrinsic
NIR polarization due to BD and YSO disks. 
This required determining the polarization properties of these foregrounds and 
backgrounds as well as developing accurate knowledge of the distances to the target disks along the line of sight so that removal of these extrinsic polarizing effects could be performed.

The Taurus star-forming region is one of the best studied laboratories for low-mass star formation 
\citep[e.g.,][]{Elias78,Beckwith90,Kenyon90} and sprawls across about $12 \times 16^\circ$ of the sky
\citep[see the Figure 18-5-6 extinction map of][]{Dobashi05}. Yet it has been
only recently, with the release of Gaia DR2 \citep{Gaia, Gaia_DR2} and the studies by \citet{Luhman18} 
and \citet{Galli19}, that accurate distances to 
the constituent dark clouds in Taurus and their associated groupings of stars, YSOs, and BDs 
have been established. 
These distances range from about 129~pc for B215 to 198~pc for L1558 \citep[both from][]{Galli19}.  
Hence, depending on
the location of the Taurus objects under study, the foregrounds and backgrounds that will contribute unrelated polarization signals
can vary considerably across the Taurus region.

Young, low-mass objects in different sub-regions of Taurus have been cataloged and investigated in previous
studies. These include the \citet[][hereafter ITG]{Itoh96} $1 \times 1$\degr\ NIR survey of the Heiles Cloud 2 (hereafter HCl2) dark molecular cloud that found 831 sources, 50 of which were classified as young (YSO Class I or II). 
Their deep, higher resolution NIR survey 
\citep{Itoh99} of 23 of the young objects revealed five of the ITG objects also had faint, nearby ($<$ 6~arcsec) low-luminosity companions. 

Multi-wavelength spectral energy distributions (SEDs) for many of the ITG objects and other young, low-mass stars
and BDs in Taurus were developed and modeled by \citet{Andrews13} to assess which had disks and envelopes and to 
ascertain many of the disk properties. Their SED fitting also refined the natures of the hosts and whether 
the systems suffered dust extinction and reddening arising outside of the host-disk system.

Portions of the Taurus system of dark clouds also have been probed to reveal magnetic field properties using background
starlight polarimetry in the NIR using the Mimir instrument \citep{Clemens07} by \citet{Chapman11} as well as 
through previously published optical and NIR studies \citep[as reviewed in][]{Chapman11}. Across much of Taurus, the predominant magnetic field orientation is perpendicular to the Galactic equator, a somewhat unusual 
condition not shared by most of the molecular material in the Galactic disk \citep{Clemens20}.

In the current study, NIR polarimetry was performed using Mimir toward two sky fields in Taurus known to contain
BDs with disks. 
Analysis of the NIR data, in combination with Gaia Early Data Release~3 \citep[EDR3:][]{Gaia, Gaia_EDR3} proper motions 
and parallaxes as well as archival NIR and mid-infrared photometry were used to ascertain foreground, embedded,
and background polarization and extinction properties. These were used to deduce the intrinsic NIR polarization
properties of the disks around the two BDs and two YSOs.

The remainder of this paper is organized as follows. The target fields and objects, the 
Mimir observations, data processing
steps, and apparent polarization properties of the objects in each field are described in Section~\ref{sec:obs}. 
Identification of foreground, embedded, and background objects and the determination of the polarization
signals contributed by the foreground and background ISM are described in Section~\ref{sec:analyses}. 
After correcting for the polarization contributions from HCl2, the resulting intrinsic polarization properties for the
BDs and YSOs are presented in Section~\ref{subsec:disks}.
Section~\ref{sec:discussion} considers the origin of the NIR intrinsic polarization are argues 
that one BD-YSO pair constitutes
a 5,200~au wide-binary with co-aligned disks. 
The past role possibly played by the magnetic field in HCl2 is assessed in light of the  
relative orientations of these disks and the local magnetic field orientation.
Finally, Section~\ref{sec:summary} provides a project summary.

\section{Target Fields and Observations} \label{sec:obs}

\subsection{Brown Dwarf and Young Stellar Object Disk Targets} \label{sec:BDs}

Analysis and modeling of archival ALMA continuum imaging, combined with other archival multi-band photometry, enabled \citet{Rilinger19} to 
improve on the \citet{Andrews13} 
models for the disks around two late-M type BDs in Taurus: ITG~17 (EPIC~248029954; CFHT~Tau~4) and IRAS~S04414+2506 (EPIC~247915927; 2M0444).  The \citet{Rilinger19} ALMA data analyses  
also partially resolved the two disks, yielding constraints on their major axis orientations and inclination angles, as well as other properties such as mass and inner and outer radii, which they sought to use to test BD 
formation models.

These two BD disk systems were selected for study using NIR polarimetry observations with the $10 \times 10$~arcmin
field of view Mimir instrument at the 1.8m~Perkins Telescope Observatory (PTO). The two non-overlapping fields 
centered on these BD targets were designated 
the ``CFHT Tau 4'' and ``2M0444'' fields, following the target names used by \citet{Rilinger19}. 
Because the Mimir imaging polarimetry solid angle is large, 
field objects (both foreground and background) for both observed Mimir regions
were simultaneously sampled for use in ascertaining the 
polarization signals contributed by the magnetic field of the intervening diffuse ISM and the denser
HCl2 in Taurus. 

The CFHT~Tau~4 field contains an additional five ITG objects (ITG~15, 16, 19, 21, and 25).
Two of these are YSOs with disks (ITG~15 and 25) that were in the \citet{Andrews13} SED analysis study.
They were added to the two BDs to become the four objects of this study.
Three of the ITG objects (15, 21, and 25) were in the close companion survey of \citet{Itoh99}. 
As summarized in Appendix~\ref{not_YSOs},
ITG~16, 19, and 21 were found to not be YSOs 
and not associated with HCl2 but instead to be unrelated background stars.
Select properties for all of the ITG objects in this field, and
those of their companions, are listed in Table~\ref{tab: prop} as are
the properties of the BD in the 2M0444 field.
Four (ITG~15, 17, 21, and 25) are doubles, identified using `A' designations for the 
brighter primaries and `B' for the fainter companions.

The 0.1~pc diameter dense core IRAS~04368+2557 (L1527) \citep{Benson89} and its embedded YSO Class~0/I source 
L1527~IRS \citep[e.g.,][]{Chen95, Motte01, Tobin08, Kristensen12} are also located within
the CFHT~Tau~4 field observed by Mimir. 
The relation of the HCl2 magnetic field orientation to this dense core, YSO, and its disk and envelope will be considered in a future paper.

\subsection{Near-Infrared Polarization Observations}

Observations of the CFHT~Tau~4 and 2M0444 fields were obtained with the Mimir instrument 
in its imaging polarimetry mode, which had a pixel field of view of $0.6 \times 0.6$~arcsec
onto a $1024 \times 1024$~pixel ALADDIN III InSb detector array,
at the PTO, located on Anderson Mesa, outside Flagstaff, AZ
on the UT nights of 2019 December 13 and 21 as well as 2020 January 6, February 12 and 14.
One polarimetric observation in the NIR $H$-band (1.6~$\mu$m) consisted of 96 images, 
obtained as single images for each of 16 distinct
orientations of the cold half-wave plate within Mimir, for each of 6 sky dither positions, offset by 
15--21 arcsec. Image integration times of 2.3, 10, and 15~sec were used for different observations.
After evaluating the images in the multiple observations for sufficiently high quality, the total useful
integration time for the CFHT~Tau~4 and 2M0444 fields was 55~min each.

Dome flats, dark current, 
and sky observations of polarization standard stars all followed usual Mimir procedures, 
as described in \citet{Clemens12a, Clemens12b}. Data processing also followed standard
Mimir steps, resulting in a catalog of polarization values
\citep[a POLCAT,][]{Clemens12c} for each observation. The cataloged values for each object included the linear Stokes
parameters $U$ and $Q$ (as percentages of Stokes $I$), the debiased polarization percentage \Pp, 
the equatorial polarization position angle (EPA, measured East from North), and the uncertainties in all these quantities. 

Polarization quantities for objects contained in the six POLCATs for the 
CFHT~Tau~4 field were combined for each matching object using Stokes $U$ and $Q$ averaging, weighted by their
variances, followed by recovery of the raw polarization percentage $P$ (= $(U^2 + Q^2)^{0.5}$) 
and its uncertainty $\sigma_P$  for 121 objects in this field. The average seeing was about 2.1~arcsec.
The same Stokes averaging was also performed for the five POLCATs obtained for the 2M0444 field, resulting in polarization information for 275 objects. The average seeing for this field was about 1.6~arcsec.

For objects with $P$ exceeding $\sigma_P$, the debiased \Pp was computed as 
$(P^2 - \sigma_P^2)^{0.5}$. For objects not meeting this criterion, \Pp was set to zero.
Equatorial polarization position angles (EPA) were computed from the Stokes parameters, and their
uncertainties were computed from the debiased polarization signal-to-noise ratio (SNR). 
Where \Pp was zero, EPA was also set to 0\degr\ and its uncertainty set to 180\degr.

Table~\ref{tab:data} presents a shortened version of the electronic table that contains the
NIR polarization values measured for the objects in the two fields. The first column offers 
R.A.-ordered field and object number designations. The next two columns provide J2000 R.A. and decl. values, 
both in degrees. The Mimir-based $H$-band magnitude is next, though users are cautioned that
no color corrections have been applied \citep[see][]{Clemens12b}. The debiased polarization
percentage \Pp and its uncertainty $\sigma_P$ are followed by the EPA
and its uncertainty. The final four columns provide the Stokes $Q$ and $U$ values and their
uncertainties. 

% Table 1 properties
\movetabledown=25mm
\begin{rotatetable}
\begin{deluxetable}{lcccccccl}
\tablecaption{Observed Fields and Selected Target Systems \label{tab: prop}}
\tabletypesize{\scriptsize}
\tablehead{\\
\colhead{Desig.}&\colhead{R.A.}&\colhead{decl.}&\colhead{Type}&\colhead{Sp. Typ.}&\colhead{Mass}&\colhead{Lumin.}&\colhead{Offset: from}&\colhead{Other Desig.}\\
&\colhead{($^\circ$)}&\colhead{($^\circ$)}&&&\colhead{($M_\odot$)}&\colhead{($L_\odot$)}&&\\
\colhead{(1)}&\colhead{(2)}&\colhead{(3)}&\colhead{(4)}&\colhead{(5)}&\colhead{(6)}&\colhead{(7)}&\colhead{(8)}&\colhead{(9)}
}
\startdata
\\
\multicolumn{9}{c}{{\it CFHT Tau 4 Field} \ \ \ \ Centered at (R.A., decl. [J2000]) = (69.95\degr,+26.02\degr) - (L, B) = (173.84\degr,$-$13.56\degr)}\\[6pt]
ITG~15B & 69.93648 & +26.03192 & ... & ... & 0.009$^{\rm a}$ & ... & 3.0-3.1$^{\prime\prime}$: ITG~15A$^{\rm a, b}$& \\
ITG~15A  & 69.93700 &+26.03131 & YSO & M5$^{\rm c}$ & 0.17$^{\rm a}$ & 0.51$^{\rm d}$ & ... & IRAS~F04366+2555,\\[-1pt]
&&&&&0.117$^{\rm b}$&0.088$^{\rm b}$&& 2MASS~J04394488+2601527\\[10pt]
ITG~16  &  69.94289 & +25.95412 & ...$^{\rm i}$ &  ... & ... & ... & ... & 2MASS~J04394629+2557149\\[10pt]
ITG~17A  & 69.94784 & +26.02797 & BD & M7$^{\rm c}$ & 0.095$^{\rm d} $& 0.175$^{\rm d}$ & 37$^{\prime\prime}$: ITG~15A & CFHT Tau 4,\\[-2pt]
&&&&&&&& 2MASS~J04394748+2601407\\
ITG~17B & 69.94899 & +26.02744 & ... & ... & ... & ... & 4.2$^{\prime\prime}$: ITG~17A$^{\rm a}$ & \\[10pt]
ITG~19 & 69.98784 & +26.09035 & ...$^{\rm i}$ & ... & ... & ... & ... & 2MASS~J04395708+2605252\\[10pt]
ITG~21B    & 70.00722 & +25.94132  & ...$^{\rm i}$ & M5.5$^{\rm c,e}$ & ... & ... & 0.56$^{\prime\prime}$: ITG~21A$^{\rm e}$ & (2MASS~J04400174+2556292)$^{\rm e}$\\
ITG~21A    & 70.00736 & +25.94141  & ...$^{\rm i}$ & M5.5$^{\rm c,e}$& ... & ... & ... & (2MASS~J04400174+2556292)$^{\rm e}$\\[10pt]
ITG~25B & 70.03213 &+26.08989 & ... & ... & 0.018$^{\rm a}$& ... & 4.3$^{\prime\prime}$: ITG~25A$^{\rm a}$ \\
ITG~25A  & 70.03335 & +26.09040 & YSO & (K6--M3.5)$^{\rm d}$ & 0.19$^{\rm a}$ & 0.90$^{\rm d}$ & ...  & IRAS 04370+2559,\\[-2pt]
&&&&M2c$^{\rm f}$&&&& 2MASS~J04400800+2605253 \\[10pt]
\multicolumn{9}{c}{{\it 2M0444 Field} \ \ \ \ Centered at (R.A., decl. [J2000]) = (71.11\degr,+25.21\degr) - (L, B) = (175.16\degr,$-$13.27\degr)} \\[6pt]
2M0444 & 71.11309 & +25.20456 & BD & M7.25$^{\rm c}$ & 0.05$^{\rm g}$ & 0.028$^{\rm h}$ & ... & IRAS~S04414+2506,\\[-2pt]
&&&&&&&& 2MASS~J04442713+2512164\\[6pt]
\enddata
\tablecomments{$^{\rm a}$~from \citet{Itoh99}, $^{\rm b}$~from \citet{WD18}, $^{\rm c}$~from \citet{Luhman10}, 
$^{\rm d}$~from \citet{Andrews13}, $^{\rm e}$~resolved by UKIDSS and Gaia DR2 and EDR3 but not by 2MASS or Mimir,
$^{\rm f}$~from \citet{Joncour17}, 
$^{\rm g}$~from \citet{Ricci13},  $^{\rm h}$~from \citet{Ricci14}, $^{\rm i}$~see Appendix~\ref{not_YSOs}}
\end{deluxetable}
\end{rotatetable}

% Table 2 data
\movetabledown=25mm
\begin{rotatetable}
\begin{deluxetable}{lccccccccccc}
\tablecaption{Observed Near-Infrared $H$-band Polarization Properties of Objects in the Fields \label{tab:data}}
\tablehead{\\
\colhead{Desig.}&\colhead{R.A.}&\colhead{decl.}&\colhead{$m_H$}&\colhead{\Pp}&\colhead{$\sigma_P$}&\colhead{EPA}&
\colhead{$\sigma_{EPA}$}&\colhead{Stokes $Q$}&\colhead{$\sigma_Q$}&\colhead{Stokes $U$}&\colhead{$\sigma_U$}\\
\colhead{(Field-Num.)}&\colhead{($^\circ$)}&\colhead{($^\circ$)}&\colhead{(mag)}&\colhead{(\%)}&\colhead{(\%)}&\colhead{(\degr)}&\colhead{(\degr)}&\colhead{(\%)}&\colhead{(\%)}&\colhead{(\%)}&\colhead{(\%)}\\
\colhead{(1)}&\colhead{(2)}&\colhead{(3)}&\colhead{(4)}&\colhead{(5)}&\colhead{(6)}&\colhead{(7)}&\colhead{(8)}&\colhead{(9)}&\colhead{(10)}&\colhead{(11)}&\colhead{(12)}
}
\startdata
% desig & ra & dec & mH & P' & sP & EPA & sPA & q & sq & u & su \\
10001\tablenotemark{a} & 69.846268  &  26.064350 & 13.822  &  2.256 & 1.317 &  71.3  & 16.7 &  $-$2.077 &  1.305 &   1.584  & 1.337 \\
10002 &  69.848568 &  26.078516 & 14.349  &  1.191 & 1.851 &  53.7  & 44.5 &  $-$0.660  & 1.871  &  2.100   &1.849 \\
10003 &  69.849897 &  26.045857 & 16.650  &  0.000 &18.765 &   0.0  & 180.0 & $-$15.536 & 18.767 & $-$0.982 & 18.383\\
... \\
20001  & 71.022033 &  25.209503 & 13.241  &  0.000 & 5.135  &  0.0  & 180.0  &  3.728 &  5.154 &   1.680 &  5.040\\
20002  &  71.024325 &   25.235553 & 13.403 & 1.501 & 0.773  & 17.4 &  14.8   & 1.385  & 0.771  &  0.965  & 0.777 \\
20003  &  71.024721 &   25.233388 & 16.361  & 0.000 &14.228 &   0.0 & 180.0  &  9.881 & 14.130 &   7.198 & 14.412 \\
\enddata
\tablecomments{Table~\ref{tab:data} is published in its entirety in the machine-readable format.
      A portion is shown here for guidance regarding its form and content.} 
\tablenotetext{a}{The initial digit identifies the R.A.-ordered field: 1 for the CFHT~Tau~4 field; 2 for the 2M0444 field. The remaining four digits encode an R.A.-ordered object serial number for each field.}
\end{deluxetable}
\end{rotatetable}

\clearpage

% Fig 1
\begin{figure}
\includegraphics[width=7.15in,trim= 0.20in 0.2in 0in 0in]{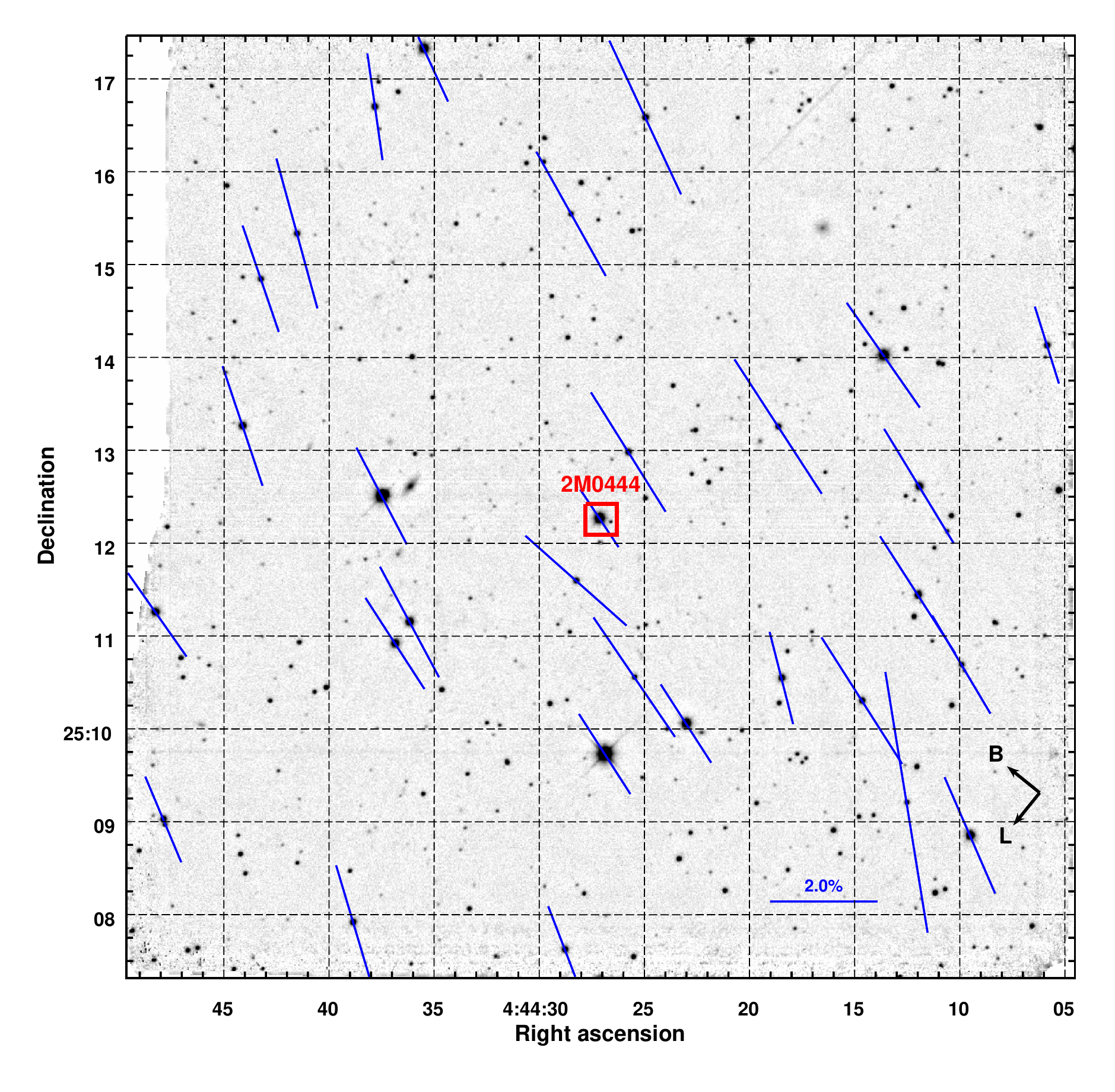}
\caption{Mimir $H$-band image of the $10 \times 10$~arcmin$^2$ field of view containing the BD 2M0444. 
Blue pseudo-vectors, with 2\% reference pseudo-vector at lower right, show
the \Pp lengths and EPA orientations for the 30 objects meeting the polarization detection criteria described in the text.
The average EPA is about 27\degr\ and the Galactic plane is parallel to EPA 141\degr, as shown via the
black B and L labeled vectors at lower right. Several galaxies appear in the image, consistent
with the average A$_V$ of 2-3~mag for the field. (Figure~\ref{fig: 2M_zoom} shows a zoomed view of the central region.)
\label{fig:2M0444}}
\end{figure}

\clearpage

\subsection{Apparent Polarization Maps} \label{sec:maps}

Figure~\ref{fig:2M0444} displays the combined Mimir $H$-band image for the 2M0444 field.
The 30 stars that met polarization quality criteria of $\sigma_P \le 1.5$\%, polarization signal-to-noise
$PSNR \equiv (P/\sigma_P) \ge 2$, and $m_H \le 16.4$~mag are 
shown as blue pseudo-vector lines (lacking ends). The lengths
of the lines are proportional to the \Pp values and their orientations show the 
EPA values. 
The central BD 2M0444 (object 20153 in Table~\ref{tab:data}) was significantly detected, 
exhibiting \Pp~$= 1.29\% \pm 0.24$\% and EPA~$= 34\degr \pm 5$\degr\
 in the $H$-band.
The unweighted average EPA of the polarization-detected objects is $26\degr.6 \pm 1.\degr7$, 
while the disk of the Milky Way has an EPA of about 141\degr. 
The similar average of \Pp values is $2.33\% \pm 0.12$\%. For the 27 polarization-detected objects that have 2MASS
\citep{Skrutskie06} matches, the average $(H - K)$ color is $0.273 \pm 0.011$~mag. Following the Near-Infrared
Color Excess (NICE) method \citep{Lada94}, the average extinction A$_V$ is 2.3~mag for those objects. 

\begin{figure}
\includegraphics[width=7.15in,trim= 0.20in 0.2in 0in 0in]{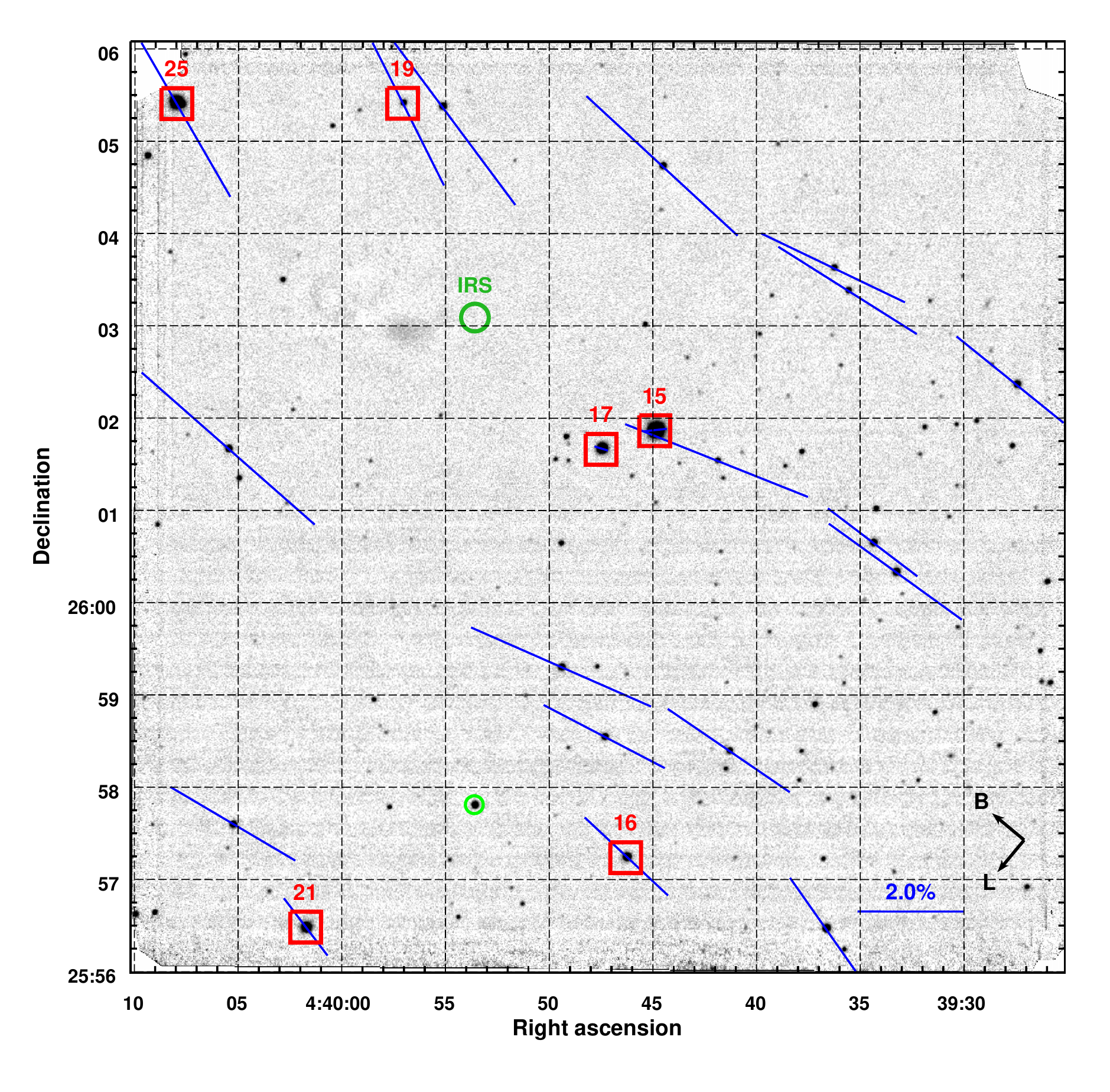}
\caption{Mimir $H$-band image of the CFHT~Tau~4 field containing the BD ITG~17 (aka CFHT~Tau~4)
and the two YSOs ITG~15 and 25. Red squares identify the 
six objects with ITG \citep{Itoh96} numbers. Blue pseudo-vectors, with 2\% reference pseudo-vector at lower right, show
the \Pp lengths and EPA orientations for the 20 objects meeting the polarization detection criteria described in the text.
One object bright enough to show polarization, but failing the $PSNR$ criterion is indicated by a light green circle at lower left (see Appendix~\ref{lowp_star}).
The location of the L1527~IRS source is indicated by a dark green circle and label at upper left.
The mean polarization EPA is 54$^\circ$, which is perpendicular to the Galactic equator orientation of 140\degr.
However, both ITG~15 and ITG~17 exhibit distinctly different polarization values than seen in the remainder
of the field. (Figure~\ref{fig: CFHT_zoom} shows a zoomed version about these two objects.)
\label{fig:CFHT4}}
\end{figure}

Figure~\ref{fig:CFHT4} shows the Mimir image for the CFHT~Tau~4 field, with the 20 polarization-detected
objects drawn as blue lines. One low-polarization object, described in Appendix~\ref{lowp_star}, is shown by a green circle.
All six ITG objects, labeled with
red boxes, had NIR values that met the three polarization selection criteria. 
The average \Pp for the 20 polarization-detected objects is $2.81\% \pm 0.27$\% at an average EPA of
$54\degr \pm 4$\degr. However, the \Pp values for the YSO ITG~15 (object 10073 in Table~\ref{tab:data}) and 
the BD ITG~17 (object 10080; CFHT~Tau~4) are far below the average, at $0.44\% \pm 0.08$\% and $0.26\% \pm 0.11$\%, respectively. 
The YSO ITG~25 (object 10117) shows a greater than average \Pp of $4.08\% \pm 0.10$\%. 
For the 20 object sample, the mean $(H - K)$ color is $0.65 \pm 0.06$~mag, 
implying an average A$_V$ of 8.2~mag for this field. 
More objects are visible in the bottom and right portions of the
Figure, while an absence of objects is noted in the upper left where one outflow lobe of L1527~IRS \citep{Cook19}
is seen as the extended faint feature East of IRS.

% Fig 2

\section{Analyses} \label{sec:analyses}

Analyses began with establishing distances to the BDs and stars that had been measured for polarization, using 
Gaia Early Data Release 3 \citep[EDR3;][]{Gaia_EDR3} parallaxes and proper motions, as described in Section~\ref{subsec:dist}. 
The properties of the extincting molecular cloud material were established via stellar reddening and molecular gas emission map comparisons, in Section~\ref{subsec:clouds}. 
Locations were determined
for the BD objects and YSOs from their extinctions, relative to the local HCl2 values, and the effects of foreground polarization on the 
BD and YSO disk polarization signals were quantified. The ITG~17 and 2M0444 BDs, as well as the YSOs 
ITG~15 and ITG~25, were found to reside 
within HCl2, as described in Section~\ref{subsec:BDs}.

In Section~\ref{subsec:UQ}, characterization and removal of the polarization signals added by the passage of the light from the BDs and YSOs through magnetically aligned dust in HCl2 revealed 
the intrinsic BD and YSO disk polarization properties (Section~\ref{subsec:disks}). These were 
compared to the properties found for the dust thermal emission from the disks detected using ALMA
by \citet{Ricci14} and \citet{Rilinger19}. 

\subsection{Cloud and Object Distances} \label{subsec:dist}

Vizier \citep{Ochsenbein00} was used to fetch Gaia EDR3, 2MASS \citep{Skrutskie06}, UKIDSS \citep{UKIDSS} 
Galactic Cluster Survey \citep{UKIDSS_GCS}, and
WISE \citep{Wright10} entries across the solid angles spanned by the two Mimir fields.
These stellar entries were position-matched to the Mimir $H$-band POLCAT entries for each field, using {\it topcat} \citep{2005ASPC..347...29T}, resulting in tables 
containing the polarimetry, multi-band photometry, parallax, and proper motion data for the objects. 

% Fig 3
\begin{figure}
\includegraphics[width=0.95\textwidth,trim= 0.15in 0.2in 0.3in 0in]{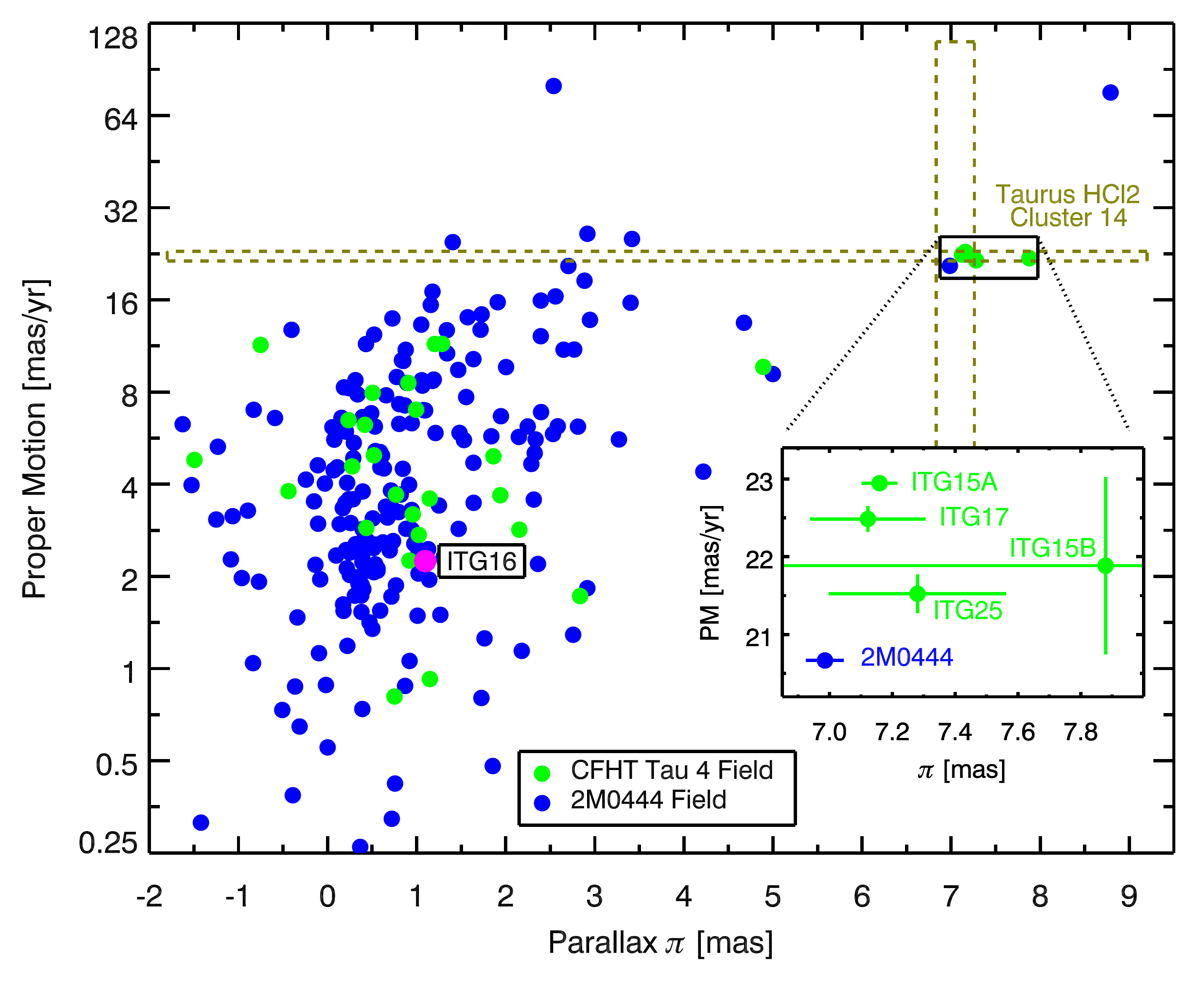}
\caption{Plot of Gaia EDR3 proper motion amplitude versus parallax for objects in the two sky fields. The 
vertical axis displays the proper motion amplitude, scaled as the base-2 log to show the full range of values.
The means and standard deviations of the values for cluster~14 of \citet{Galli19} in Taurus HCl2 
are indicated by the vertical and horizontal olive-green dashed rectangles. Four of the CFHT~Tau~4 field ITG objects and 2M0444 have Gaia EDR3 matches and are labeled with the same color as their field. 
No objects in the CFHT~Tau~4 field (filled green circles) are located closer than the indicated ITG objects. One object in the 2M0444 field (filled blue circles) is closer
than the BD in that field. 
ITG~16 in the CFHT~Tau~4 field, labeled next to its magenta filled circle, has Gaia EDR3 parallax and 
proper motion values placing it far more distant than HCl2.
{\it Inset}: Expanded view of proper motion and parallax for the objects inside the boxed region shown near
the cluster~14 label. Error bars shown indicate Gaia~EDR3 $\pm$1~$\sigma$ uncertainties. 
\label{fig: P_PM}}
\end{figure}

\citet{Luhman18} and \citet{Galli19} analyzed the Gaia DR2 parallaxes and proper motions over much larger 
portions of the Taurus cloud complex, establishing accurate distances to the individual molecular clouds 
making up the Taurus complex and finding groups of objects exhibiting similar distances and proper motions. 
\citet{Galli19} used cluster analysis techniques to identify more distinct star groups than \citet{Luhman18} 
by also including spatial clustering. 
Of the 21 clusters \citet{Galli19} identified and characterized (but noting that these could be unbound groupings of objects that merely exhibit similar distances, motions, and spatial locations), their cluster~14 best matches the BD and YSOs in the CFHT~Tau~4 field and the 2M0444 BD. 
Indeed, ITG 15A and 17A as well as 2M0444 are contained in the cluster~14 inventory reported in Appendix~A of \citet{Galli19}, while the remaining YSO, ITG 25A, they assigned to their cluster~15. 

Figure~\ref{fig: P_PM} shows the amplitude of proper motion versus parallax for 
objects in the two Mimir fields that had such information listed in Gaia~EDR3.
The means and standard deviations of the amplitude of proper motion and the
parallax for cluster~14 from \citet{Galli19} are shown as the vertical and horizontal olive-green 
dashed rectangles that pass through the groupings of ITG objects plus 2M0444. 
The Gaia measured uncertainties in proper motions and parallaxes depend strongly on brightness,
ranging from 0.045 to 1.76~mas~yr$^{-1}$ for 2M0444 and ITG15B, respectively, and 0.057 to 
1.20~mas for ITG~15A and ITG~15B, respectively. 
One finding from Figure~\ref{fig: P_PM} is that to the limits of Gaia~EDR3, no 
significant foreground stellar population was revealed. The improved
Gaia~EDR3 values, displayed in the Figure~\ref{fig: P_PM} inset, tighten the proper motion and parallax
ranges with respect to the Gaia~DR2 values, confirming that the four ITG objects are part of the same association in Taurus at a distance of 139.6$\pm$1.0~pc, computed as the inverse of the weighted parallax mean and propagated uncertainty. The distance to 2M0444 is similar, though greater by 3.6$\pm$1.6~pc.

\subsection{Cloud Structures and Extinctions} \label{subsec:clouds}

Extinction maps by \citet{Dobashi05} and \citet{Lombardi10} have revealed the distribution of dust in 
Taurus, while the molecular gas has been 
traced by \citet{FCRAO} using CO and $^{13}$CO. 
Here, the molecular gas distribution
in each field was obtained from the \citet{FCRAO} data as a line-integrated $^{13}$CO spectral line map by 
limiting the radial velocity integration window to be from 0 to 12~km~s$^{-1}$. 
This gas tracer is expected to faithfully reveal gas column densities, provided
the gas is not too diffuse, so that molecular self-shielding fails and dissociation dominates, and provided the the gas
is not too volumetrically dense, so that CO is depleted as mantles onto dust grains \citep[c.f.,][]{Pineda10}. 
In HCl2, both of these processes
likely occur, so the molecular maps will be only partially representative of the gas properties along these lines of sight.

% Fig 4
\begin{figure}
\includegraphics[width=7.00in,trim= 0.1in 0.35in 0in 0.15in]{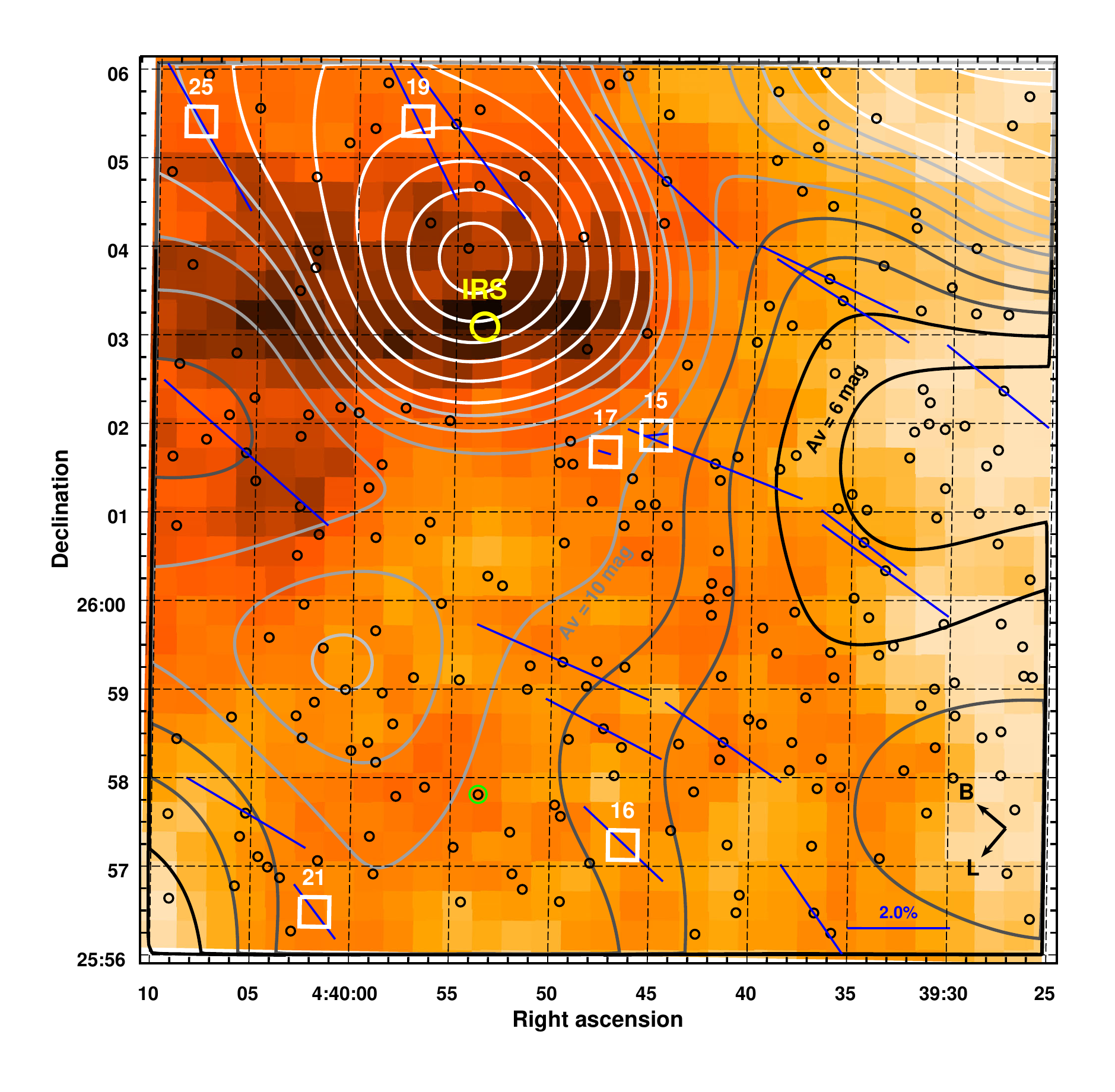}
\caption{Map of $^{13}$CO line-integrated emission and extinction A$_V$ across the CFHT~Tau~4 field. 
The color image shows $^{13}$CO integrated emission, from \citet{FCRAO}, ranging from about 2.3~K~km~s$^{-1}$
for the lightest yellow colors up to about 5.1~K~km~s$^{-1}$ for the darkest colors, at about 45~arcsec resolution.
The contours indicate inferred A$_V$ values, at about 2~arcmin resolution, derived from the $(H - M)$ colors of the 
objects but not including the ITG objects.
Contours trace 6 to 13~mag of A$_V$ in steps of 1~mag, followed by 14 to 24~mag in steps of 2~mag. 
The lowest extinctions are shown by black contours,
intermediate extinctions by two levels of gray contours, and the highest extinctions by white contours. 
The positions of the objects used
to generate the A$_V$ image are indicated by black circles. The ITG objects are shown by white boxes and labels. The location
of the Class~0/I source L1527~IRS is shown as a yellow labeled circle. 
ITG~15 and ITG~17 are located near the A$_V = 10$~mag contour,
while ITG~25 is in a region of greater extinction.
\label{fig: CFHT_Av}}
\end{figure}

The color $(H - M)$ has been shown by \citet[][i.e., the Rayleigh Jeans Color Excess method]{Majewski11} to trace dust reddening to objects and to suffer much less sensitivity to differences among the intrinsic colors of the objects as compared to the $(H - K)$ based RICE \citep{Lada94} method.
The deep Mimir observations, augmented
by UKIDSS \citep{UKIDSS} observations, when combined with WISE \citep{Wright10} band~2 
(W2; 4.62~$\mu$m $\sim M$-band), provide suitable line-of-sight $(H - M)$
reddening probes for inferring dust extinction maps \citep{Clemens16}.  In the CFHT~Tau~4 field, Mimir $H$-band and
WISE W2-band returned 176 matches. To augment this, UKIDSS data from the Galactic Cluster Survey 
\citep[GCS,][]{UKIDSS_GCS} were also compared to WISE W2. However, the GCS in this region contains no $H$-band data and
instead reports $K$-band magnitudes. 

To increase the stellar sampling through utilizing the UKIDSS $K$-band magnitudes, 
the $(K - M)$ colors of objects with $(H - M)$ colors were compared to ascertain a suitable conversion. 
Considering all such matched objects in both fields, a conversion
relation of $(H - M) = 0.15 + 1.59\  (K - M)$ was determined and adopted, though there are minor field-to-field differences likely
due to dust grain growth. For both the conversion determination and for the final
$(H - M)$ map, values for the ITG objects plus 2M0444 were excluded. 
This was done to protect the desired A$_V$ map from the effects
of intrinsic source reddening due to infrared excesses from the disks (and perhaps envelopes) around some of these objects. 
The remaining objects are expected to be mostly distant, older main sequence stars and giants without circumstellar dust 
and intrinsic reddening. 
The final set of objects with measured, or extrapolated, $(H - M)$ values in the CFHT~Tau~4 field numbered 226. 
An interpolation program generated an A$_V$ map at 5~arcsec
sampling across the field using gaussian weighting by offset as well as weighting by color uncertainties. 
The resulting map was smoothed to an effective angular resolution of about 2~arcmin so that
about 10 objects had their colors sampled for computing each resolution element.

Figure~\ref{fig: CFHT_Av} displays, for the CFHT~Tau~4 field, the integrated intensity of $^{13}$CO as the color background image. Overlaid
are contours of dust extinction, A$_V$, estimated as $7.6 \times (H - M - 0.08)$, which represents normal dust ($R_V \sim 3.1$)
but underestimates extinction traced by larger dust grains, which tend to be grayer ($R_V \sim 5$). The black 
circles in the Figure identify the locations of the objects used to derive the A$_V$ contours. Note, again, that the ITG objects
were not included, as their colors are to be used to estimate their locations with respect to the extincting, and 
presumably polarizing, dust. Both the $^{13}$CO and dust show that the highest column densities are associated
with the deeply embedded Class~0/I L1527~IRS source at upper left.

% Fig 5
\begin{figure}
\includegraphics[width=7.00in,trim= 0.1in 0.35in 0in 0.05in]{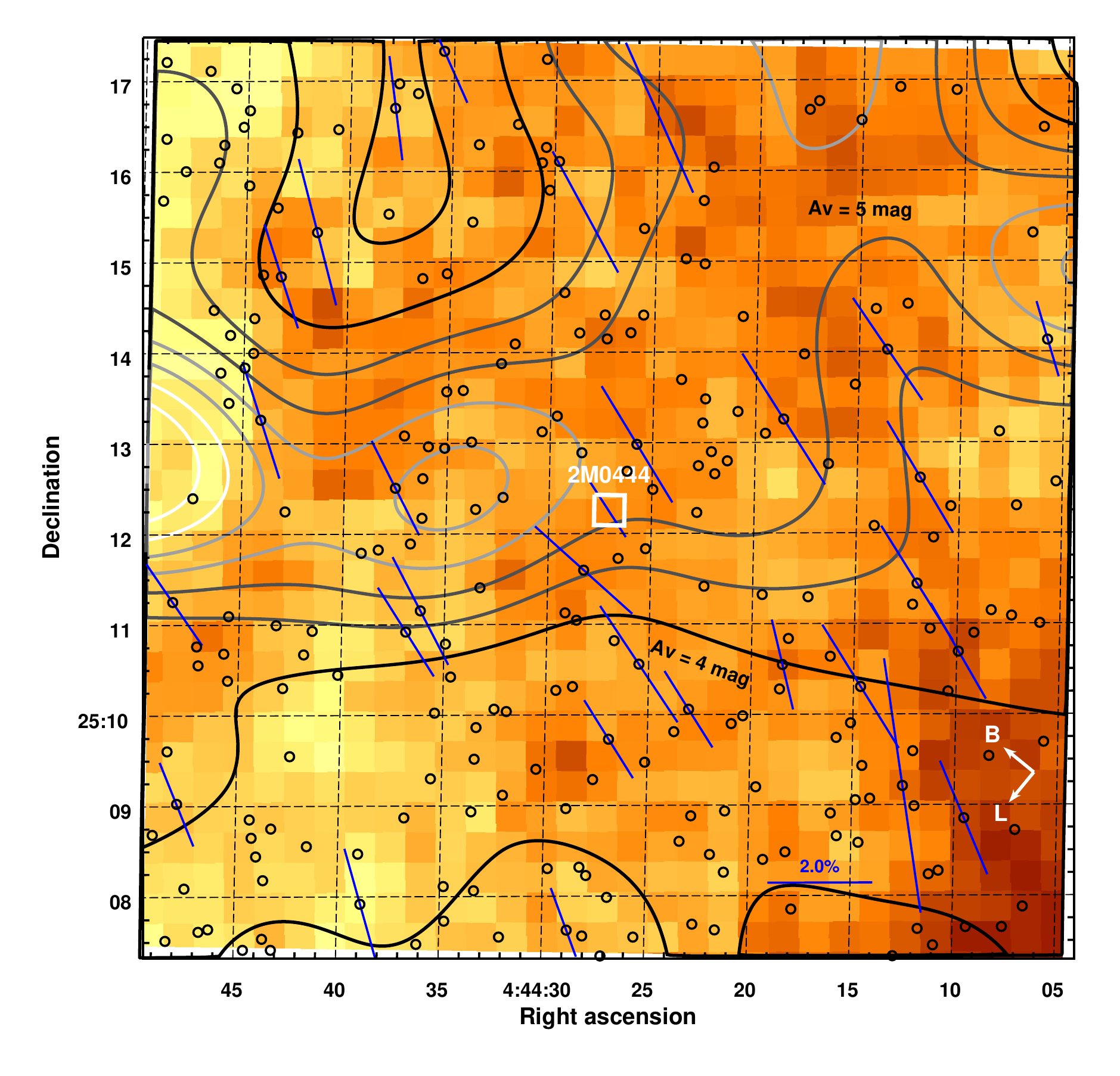}
\caption{Map of $^{13}$CO line-integrated emission and extinction A$_V$ across the 2M0444 field.
The color image shows $^{13}$CO integrated emission, from \citet{FCRAO}, ranging from about 0.3~K~km~s$^{-1}$
for the lightest yellow colors up to about 2.2~K~km~s$^{-1}$ for the darkest colors, near the bottom right corner.
The contours indicate inferred A$_V$ values, at about 2~arcmin resolution.
Contours trace 3.5 to 7~mag of A$_V$, in 0.5~mag steps. Lowest extinctions are shown by the black contours,
intermediate by gray contours, and highest by the white contours. The positions of the objects used
to generate the A$_V$ image are indicated by black circles. The 2M0444 BD is shown by the labeled white box.
\label{fig: 2M_Av}}
\end{figure}

A similar map for the 2M0444 field is shown as Figure~\ref{fig: 2M_Av}. 
The background color image is the integrated intensity of $^{13}$CO, displayed from 0.3 to 2.2~K~km~s$^{-1}$.
The contours indicate levels of extinction A$_V$, from 3.5 to 7~mag. The CO gas and the extincting
dust show some correlation, but not as strongly as for the CFHT~Tau~4 field. 
The open
black circles show the locations of the 233 objects sampled for reddening, with 195 yielding $(H - M)$
colors from the Mimir observations and WISE and another 38 coming from $(K - M)$ from UKIDSS and
WISE, suitably scaled as was done for the CFHT~Tau~4 field. The BD 2M0444 was excluded from 
the A$_V$ determinations.

Extinction along the direction to 2M0444 is of the order of A$_V \sim 5$~mag, based on the extinctions
shown by objects in the vicinity that exhibit 
similar polarization properties. The deduced value of zero mag by \citet{Bouy08}, and cited by
\citet{Rilinger19}, would have a low probability of producing the polarization seen for 2M0444,
so that null extinction value is judged to be suspect.

\subsection{Brown Dwarf and YSO Locations} \label{subsec:BDs}

In Figure~\ref{fig: CFHT_Av}, the ITG~15 and ITG~17 systems are located on the sky
about halfway between the highest column density peak and regions of least extinction in the field.
If this YSO and BD were located in front of the HCl2 dust cloud, they should exhibit 
extinctions (corrected for their disk NIR excesses) less than those predicted
for their positions, which are based on the background objects probing the dust layer. 
If the ITG~15 and ITG~17 systems are located within the HCl2 dust cloud, their extinctions
should be comparable to, but perhaps not quite as great as, the background star predicted values. 
Finally, if the YSO and BD 
are far behind HCl2, their extinctions should be close to the
predicted extinction values for the dust as probed by the background objects.

However, the observed $(H - M)$ colors of the BDs and YSOs are combinations of  
foreground extinction and the infrared excess in their SEDs, primarily produced by their
disks \citep[e.g.,][also see Figure~\ref{fig: SED} in Appendix~\ref{subsec:SEDs} here]{Andrews13}.
For the objects that had their SEDs modeled in enough detail to separate these two reddening components,
namely the BDs ITG~17 and 2M0444 and the YSOs ITG~15 and ITG~25,
comparison to the cloud extinctions is straightforward. The objects that lack such detailed SED modeling
represent an additional challenge, as followed in Appendix~\ref{not_YSOs}.

Table~\ref{tab:Av} summarizes, for the two BDs and the two YSOs, 
the observed (A$_{V_O}$), interpolated (A$_{V_I}$), and modeled (A$_{V_M}$) values of A$_V$ and 
introduces two ratios to aid location determinations.
Column (2) lists the A$_{V_I}$ values found to be present along the directions to
the target objects in the interpolated A$_V$ images, shown as Figure~\ref{fig: CFHT_Av} and 
Figure~\ref{fig: 2M_Av}.
The A$_{V_O}$ values estimated directly from the observed $(H - M)$ values for the objects are listed in column~(3), 
where $H$-band is from the Mimir observations and the $M$-band is from WISE W2-band photometry.
The $^{13}$CO integrated intensity along the direction including each listed object is presented in 
column~(4). While there is a weak correlation of this quantity with the interpolated A$_{V_I}$ values, 
the dynamic range of the $^{13}$CO is limited, likely as a result of
a combination of optical depth effects and freeze-out of that molecular tracer onto
dust grains for the greater extinction lines of sight.
The A$_{V_M}$ values obtained through SED modeling of the four objects 
by \citet{Andrews13}, \citet{Bouy08}, \citet{Zhang18}, and \citet{WD18} are listed in column~(5). 
These values are in good agreement for ITG~15 and ITG~17 but differ for ITG~25 and 2M0444. 
In the following analyses, the \citet{Andrews13} A$_{V_M}$ values were adopted. 

Two ratios of A$_V$ values were created, as listed in columns~(6) and (7) of Table~\ref{tab:Av}. 
The first, $R_1$, is the ratio 
of the SED-modeled A$_{V_M}$ for an object (column~(5)) to the locally interpolated
A$_{V_I}$ value through HCl2 (column~(2)) along the same direction. The second, $R_2$, is the ratio of the 
observed $(H - M)$-based A$_{V_O}$ 
value for the object (column (3)) to the interpolated A$_{V_I}$ value (column (2)). 

% Table 3 Av values
\begin{deluxetable}{lcccccc}
\tablecaption{Dust Extinctions - Interpolated, Observed, and Modeled \label{tab:Av}}
\tablewidth{7.5truein}
\tablehead{\\
\colhead{Desig.}&\colhead{A$_{V_I}$}&\colhead{A$_{V_O}$}&\colhead{$^{13}$CO Integrated}&\colhead{A$_{V_M}$}&\colhead{$R_1$}&\colhead{$R_2$}\\
&\colhead{Interpolated}&\colhead{Observed}&\colhead{Intensity}&\colhead{Modeled}&\colhead{((5)/(2))}&
\colhead{((3)/(2))}\\
&\colhead{(mag)}&\colhead{(mag)}&\colhead{(K~km~s$^{-1}$)}&\colhead{(mag)}&\\
\colhead{(1)}&\colhead{(2)}&\colhead{(3)}&\colhead{(4)}&\colhead{(5)}&\colhead{(6)}&\colhead{(7)}
}
\startdata
ITG~15 & 10.1  &  11.8 & 3.5  &  4.16$\pm$0.54$^{\rm a}$ & 0.41$\pm$0.05 &  1.17$\pm$0.04  \\[-3pt]
&&&&4.35$\pm$0.85$^{\rm b}$\\[-3pt]
&&&&0.5$^{\rm c}$\\[3pt]
ITG~17 &  10.6 & 16.2 & 3.8  & 5.67$\pm$0.89$^{\rm a} $& 0.53$\pm$0.08 &  1.53$\pm$0.04  \\[-3pt]
&&&&6.37$\pm$0.85$^{\rm b}$\\[3pt]
ITG~25 & 13.0 & 23.8  & 4.1 & 10.65$\pm$0.75$^{\rm a}$  & 0.82$\pm$0.06 & 1.83$\pm$0.03  \\[-3pt]
&&&&2.6$\pm$4.0$^{\rm b}$\\[10pt]
2M0444 &  5.1 &   11.3 & 1.3 & 2.05$\pm$1.12$^{\rm a}$ & 0.40$\pm$0.22 & 2.22$\pm$0.09 \\[-3pt]
&&&& 0.0$^{\rm d}$& 0.0 & \\
\enddata
\tablecomments{Uncertainties are about 0.1~mag for column (2), 0.4~mag for column (3), and 0.1~K~km~s$^{-1}$ for
column (4). References: $^{\rm a}$~\citet{Andrews13}, $^{\rm b}$~\citet{Zhang18}, $^{\rm c}$~\citet{WD18},  $^{\rm d}$~\citet{Bouy08}}
\end{deluxetable}

If all of the objects providing $(H - M)$ values were located behind the extincting and polarizing dust material associated
with HCl2 and none of those objects exhibited infrared excess emission, then the corresponding $R_2$ values 
computed for those objects should all be near unity. 
Some spread of $R_2$ values could result due to extinction variations across the fields of view being only 
partially captured by the 2~arcmin resolution of the interpolations used. 
The $R_2$ distribution for each field was developed from the $(H - M)$ values for all objects indicated in Figures~\ref{fig: CFHT_Av} and \ref{fig: 2M_Av}
as black circles, again holding aside the ITG and 2M0444 objects. 
For each $(H - M)$ value, the propagated uncertainties were used to create gaussian probability distributions that
were normalized and accumulated. This yielded more representative net probability distributions, though
low SNR objects contribute distribution broadening. 

\begin{figure}
\includegraphics[width=\textwidth,trim= 0in 1.0in 0in 0in,clip]{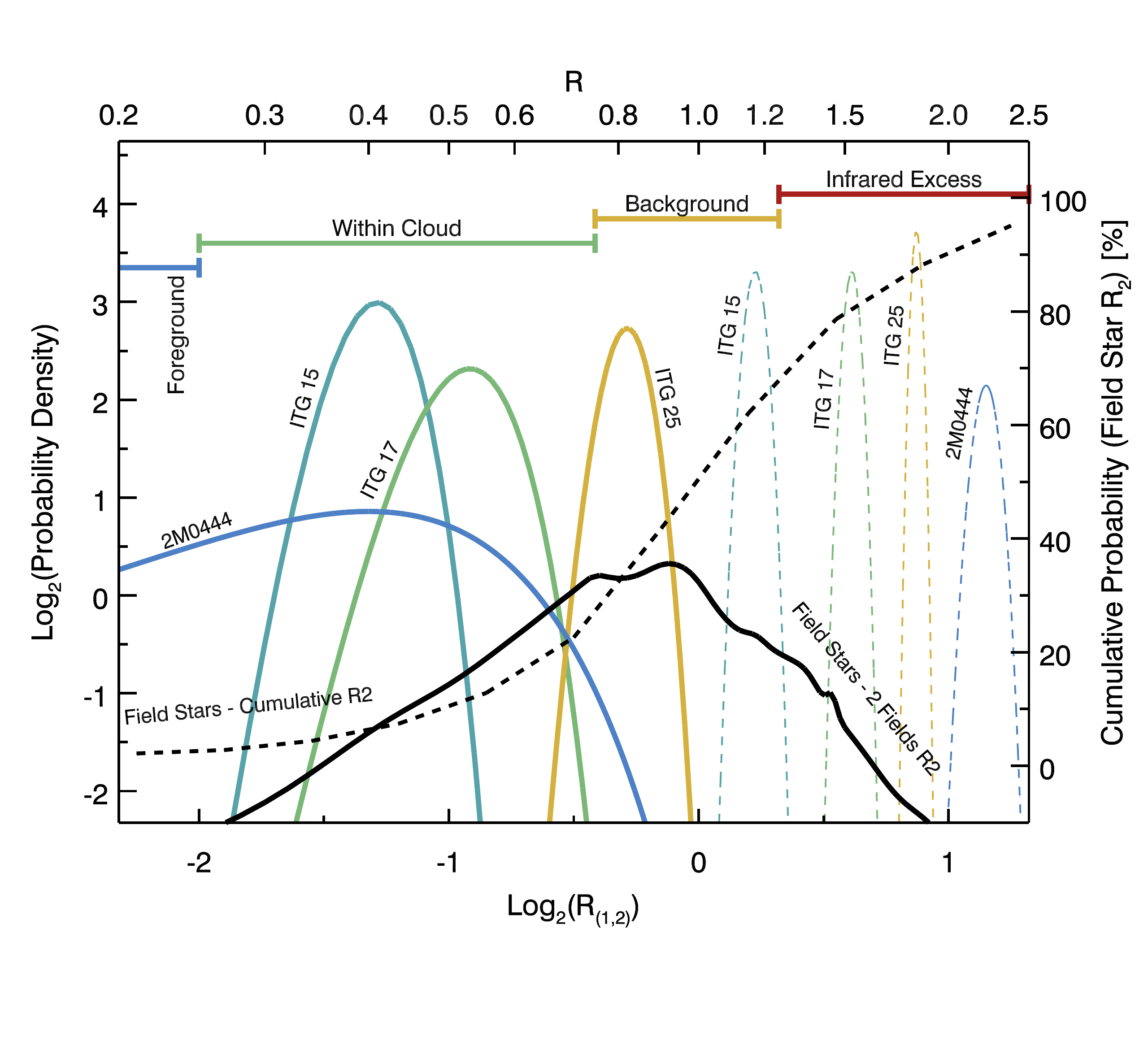}
\caption{Probability distribution functions for A$_V$ ratio functions $R_1$ {\bf ($\equiv$ A$_{V_M}$ / A$_{V_I}$)} and 
$R_2$ {\bf ($\equiv$ A$_{V_O}$ / A$_{V_I}$)}. Horizontal
axis displays the $R$ value, as base-2 log values. Left side vertical axis is 
the base-2 log of the probability density. Right side scale is linear cumulative probability
percentage. The solid black curve displays the $R_2$ probability density for the background objects in the two observed
fields, as described in the text. The dashed black curve is the corresponding cumulative $R_2$
probability. Solid colored curves display $R_1$ probability densities for 2M0444 (blue), ITG~15 (teal), ITG~17 (green),
and ITG~25 (yellow). Dashed colored curves display $R_2$ probability densities. 
Labeled horizontal colored lines across the top indicate approximate location
predictions with respect to HCl2. 
All four of the target objects are judged to be embedded within HCl2.
\label{fig: R1R2}}
\end{figure}

The resulting distributions for the
two fields were similar enough that the area normalized distributions were averaged and are plotted as the
solid black curve in Figure~\ref{fig: R1R2}. There, the left side vertical axis shows the 
probability density, as the base-2 log, and the horizontal axis represents $R$ values (both $R_1$ and
$R_2$ are plotted using the same horizontal axis), also in base-2 log form.
The dashed black line displays the cumulative probability for the $R_2$ values for the two fields, referenced to the right side
linear vertical axis. The 50\% cumulative probability for objects in both fields occurs at an $R_2$ value of 0.98, with 
16\% and 84\% cumulative probabilities found at $R_2$ values of 0.62 and 1.65. 

% Fig 6

For the set of background objects used to create the interpolated A$_V$ images and contours, 
the integrated $R_2$ likelihood below
0.25 is less than 5\%. Values of $R_2$ this low, or lower, would be expected for nearly all bona fide foreground
objects, as the extinction in the diffuse ISM foreground to HCl2 is expected to be nearly negligible. A value for this 
inner cloud boundary of less than 0.25 could have been chosen instead, but that might have caused foreground stars to be misclassified
as embedded due to photometric uncertainties or small-scale cloud structure variations.

Objects
embedded within HCl2 would not be expected to suffer the full line-of-sight extinction seen to bona fide 
background objects. Objects with $R_2$ values greater than the 0.25 inner boundary but still 
less than about unity could be deemed embedded. A rough outer cloud limit of 0.75 for $R_2$ was judged
to represent a fair compromise that accommodated expected cloud structure and photometric variations in a spirit similar
to the choice for the inner boundary. Objects with $R_2$ values well beyond unity are surely background
to HCl2 with one significant exception. Objects with disks or envelopes have SEDs\footnote{SEDs for the four target objects as well as others in the field are shown in Figure~\ref{fig: SED}.} featuring emission from
those dusty components in excess of that expected of their central object photospheres. These infrared excesses
are a hallmark of BDs and YSOs with disks (and/or envelopes) and objects showing such excesses should not 
immediately be interpreted under the $R_2$ formalism as being located behind HCl2. 

Figure~\ref{fig: R1R2} displays the $R_1$ and $R_2$ information for the three ITG objects and 2M0444 
as solid probability density curves for their $R_1$ values and as dashed curves for their $R_2$ values. Labeled solid
lines across the top of the Figure indicate the approximate location assignment zones. 

The $R_1$ curves for ITG~15, ITG~17, and 2M0444 lead to their classifications as being embedded within HCl2. 
The ITG~25 SED-modeled
extinction places this object on the boundary between being embedded and being considered background. However,
it is unlikely to be far in the background, given that its parallax and proper motions match the other
objects in cluster~14 of HCl2 \citep[][and Figure~\ref{fig: P_PM} here]{Galli19}. Note that the $R_2$ (dashed colored) curves, which ignore the SED-apportionment of A$_V$, for these same four objects in Figure~\ref{fig: R1R2} all exhibit infrared excesses, as was expected for these disk-containing systems. 

While there is a small chance that ITG~15 and ITG~17 are located just in front of HCl2 and
so do not partake of the polarization impressed on starlight by HCl2, it is far more likely that
both systems are embedded within HCl2 and that their light is affected 
by the magnetic field and dust in HCl2. The SED modeling by both \citet{Andrews13} and
\citet{Zhang18} return
A$_{V_M}$ values of 4--6 for these objects, effectively excluding foreground or cloud front surface locations.

\subsection{Magnetic Field Polarization Properties of Heiles Cloud 2} \label{subsec:UQ}

Having established that the BDs ITG~17 and 2M0444, along with the YSOs ITG~15 and ITG~25, are
embedded within HCl2, the next step is to determine the polarization properties
impressed on their starlight by HCl2. Doing so will allow those properties to be removed from the measured
polarization signals of these target objects and so reveal the intrinsic polarization properties associated
with the BD and YSO disks.

Gaia distances alone cannot provide the information sought, as the front and back locations of 
HCl2 along the directions sampled cannot be resolved with the angular and distance 
resolutions needed. Instead, a process that uses the relative extinctions developed in the
previous section was developed and applied.

\subsubsection{Stokes $U$ and $Q$ maps}\label{sec:UQ_maps}

Maps of the observed Stokes parameters were created to allow characterizing and
correcting for the HCl2 polarization contributions. The methodology
was the same as used to create Figures~\ref{fig: CFHT_Av} and \ref{fig: 2M_Av}, 
from the $(H - M)$ values.
Described in \cite{Clemens16}, the software forms interpolated maps of Stokes~$U$ and
Stokes~$Q$ using all of the measured values, applying gaussian weighting by offset from
each object to each map pixel and variance weighting by Stokes parameter uncertainties. Because of the
gaussian nature of the Stokes parameters and the variance weighting, using {\it all} 
of the observed values, including
polarization upper limits, increases information and angular resolution with little increased
noise. The BDs and YSOs were excluded from
the interpolation input data sets, so that correction by the resulting Stokes maps, using
values interpolated to the positions of the BDs and YSOs, could be performed as
was done for the A$_V$ maps.

In the CFHT~Tau~4 field, 118 objects that were not the designated targets
had Stokes parameter information measured
using Mimir in the $H$-band. For a $10 \times 10$~arcsec grid of synthetic pixel positions across the observed 
Mimir field,  all non-target objects within 4~arcmin offset were sampled for their Stokes $U$ and $Q$ values, with weighting by
the variance of those quantities and weighting by a gaussian of offset of each object from
the grid position. For the CFHT~Tau~4 field, an offset weighting gaussian with full width at 
half maximum (FWHM) of 4~arcmin provided adequate
numbers of objects for each grid position and yielded good SNR for the predicted 
Stokes parameters at the positions of the ITG objects. Smaller FWHM values resulted in poorer
map coverage and higher predicted uncertainties for the Stokes parameters. 
Greater FWHM values would
result in smaller uncertainties in the predicted Stokes values, with use of the all of the data
to compute a full-field weighted average representing the extreme application of this
approach. But, in Figure~\ref{fig:CFHT4}, there appear to be bona fide differences in the polarization properties
of HCl2 across the CFHT~Tau~4 field that need to be included in the Stokes accounting
if accurate values for the BD and YSO disk polarizations were to be achieved, hence
4~arcmin seemed to be a useful compromise resolution.

% Fig 7
\begin{figure}
\includegraphics[width=\textwidth,trim= 0.3in 0.35in 0.2in 0.0in]{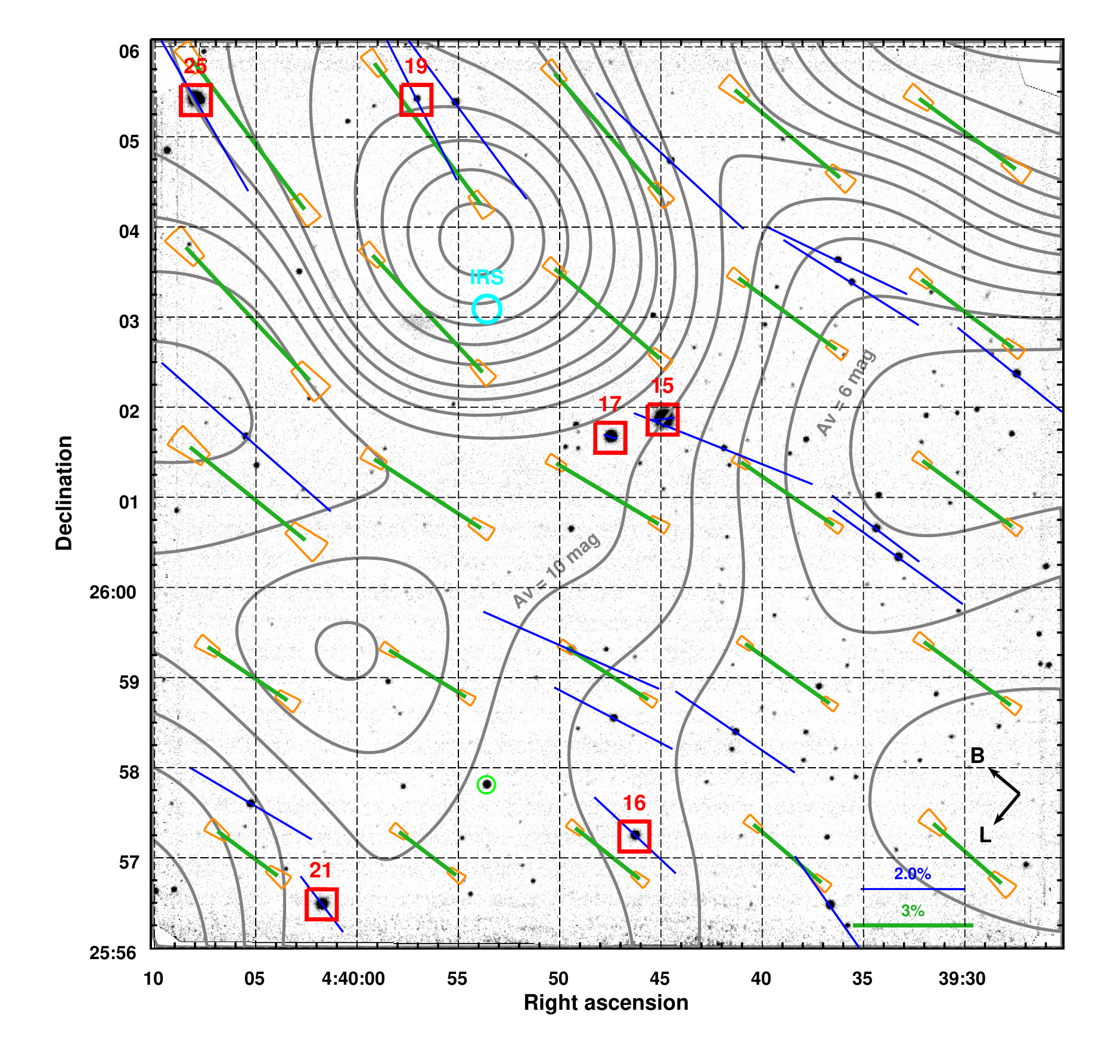}
\caption{Synthetic polarization pattern derived from the HCl2 Stokes $U$ and $Q$ interpolations
for the CFHT~Tau~4 field. The background Mimir $H$-band image, blue lines, and red boxes are the 
same as in Figure~\ref{fig:CFHT4}. Contours and contour labels are the same as in 
Figure~\ref{fig: CFHT_Av}, though all contours are colored gray here. Green pseudo-vector lines display
interpolated $P^\prime_I$ and EPA$_I$ values on a Nyquist-sampled grid, spaced at half of the 4~arcmin
FWHM resolution. Orange torus segments at the ends of the green 
lines indicate $\pm$2$\sigma$ uncertainties in $P^\prime_I$ and EPA$_I$, as described in the
text. Significant changes in $P^\prime_I$ and EPA$_I$ are evident across the field as the 
changing lengths and orientations of the green lines.
\label{fig: CFHT_Pol}}
\end{figure}

Figure~\ref{fig: CFHT_Pol} presents synthetic polarization information for the CFHT~Tau~4
field, computed from the interpolated Stokes $U$ and $Q$ arrays as a Nyquist-sampled
($2 \times 2$~arcmin) grid of values. The resulting 25
pseudo-vectors display the interpolated debiased polarization percentage $P^\prime_I$ and 
equatorial polarization position angle EPA$_I$ via their lengths and orientations. 
The $\pm 2$~$\sigma$ uncertainties for $P^\prime_I$ and EPA$_I$ are encoded in the orange torus
segments at the ends of each green pseudo-vector. Thus, for each synthetic pseudo-vector,
there is a 90\% likelihood that the pseudo-vector ends are constrained to lie within the orange uncertainty
torus segments.

For the 2M0444
field, there were 275 non-target objects available, more than twice as many as in the CFHT~Tau~4
field, enabling use of FWHM resolutions smaller than 4~arcmin.
In this field, 2~arcmin still produced low SNR values and 4~arcmin missed some of the smaller-scale
changes revealed at 3~arcmin, so this latter value was chosen as the best value. 

\subsection{Foreground Stokes Parameters Correction Methods}\label{subsec:methods}

Correcting the observed target polarization values by the interpolated Stokes parameters
across each Mimir field could follow any of three methods.
The first, and simplest, approach would be to directly subtract the interpolated
Stokes parameters at the positions of the target stars from the observed Stokes
parameters for each target, to obtain residuals that represent the intrinsic NIR polarization of the
BD or YSO systems. The drawback to this bulk correction approach is that the contaminating 
polarization signal contributed by HCl2 might depend on the depth of the target
systems within the cloud. A YSO on the back side of the cloud would suffer nearly
all of the contamination provided by the intervening cloud while a YSO close to 
the near side of the cloud would not. Hence, a second approach would be to scale the
background starlight determined Stokes parameters by the extinction
to the target system relative to the full extinction through the cloud in that direction.
This scaling factor would be the $R_1$ ratio described earlier. However, this
approach, though an improvement over the simpler one, fails to recognize that
polarization and extinction contributions are not linearly related in dark molecular clouds.

The third approach adopted here is based on a determination of how polarization
scales with extinction for each of the observed fields and extends that to an
illuminated slab model to estimate the best apportioned Stokes parameters
to use as foreground corrections, as described in the following section.

Alternate approaches also exist that would take into account the clumpy,
turbulent, 
or fractal nature of molecular clouds and might also account for small-scale local cloud
core density enhancements, usually associated with star forming regions but perhaps still 
present here after these BDs and YSOs moved into their Class~II phases.
Unfortunately, none of these methods have any background stellar 
polarimetry observations to guide or constrain them, leaving them in the
realm of speculation for now.

\subsubsection{Polarization Efficiency Dependence on Extinction}\label{subsec:PE}

The dust grains responsible for extinction and
polarization are generally considered well-mixed, if not identical. But, were this strictly
the case, observed polarization percentages $P^\prime_O$ should rise linearly with observed
extinctions A$_{V_O}$.
Such linear behavior would cause polarization efficiency (PE $\equiv P^\prime_O$/A$_{V_O}$) 
to be independent of A$_{V_O}$.
An opposite view is that dust grains align to magnetic fields only in the outer portions
of molecular clouds and either the magnetic fields are excluded from cloud interiors or dust grains
do not align to any magnetic fields located there \citep[e.g.,][]{Goodman95,Arce98}. In this case, PE is high at 
low extinctions but should decrease
as A$_{V_O}^{-1}$ for higher extinctions. This would yield a power law index of
$-1$ in a PE versus A$_{V_O}$ plot. 

In practice, neither flat nor $-1$ slopes are seen, with observed indices ranging from
about $-0.5$ to $-0.9$ \citep{Andersson15,Pattle19}. This has been interpreted to indicate that
although dust grain alignment efficiency does decrease with optical depth into molecular 
clouds, magnetic fields are still detected via dust grain alignment, though with less alignment efficiency at higher
extinctions. This has been most readily explained by the micro-physical model of radiative aligned torques 
\citep[RATS; c.f.,][]{Lazarian07,Andersson15} for dust grains.

% Fig 8
\begin{figure}
\includegraphics[width=\textwidth,trim= 0.2in 0.30in 0.1in 0.2in]{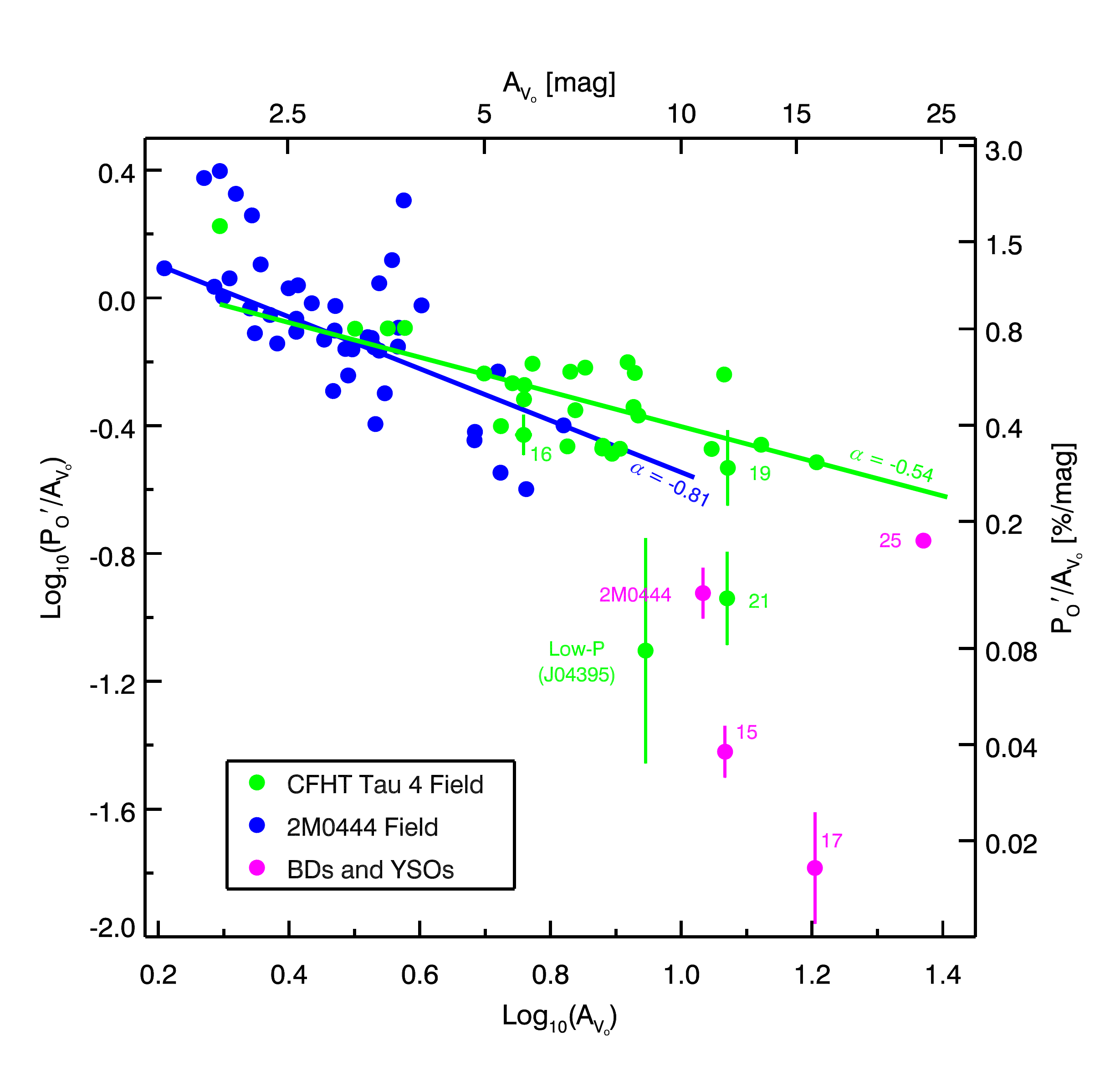}
\caption{Comparison of apparent (observed) polarization efficiency (PE = $P^\prime_O$ / A$_{V_O}$) versus 
extinction A$_{V_O}$ for objects in the two fields, in log-log form. 
Blue points represent objects in the 2M0444 field and green 
points are for objects in the CFHT~Tau~4 field. Lines in those same colors represent power
law fits, with indices $\alpha$ shown near each line. Both show slopes that are shallower than
$-1$ and steeper than zero. Magenta points are the three ITG objects plus 2M0444. 
The green
numbered ITG stars are discussed in Appendix~\ref{not_YSOs}.
The 
green filled circle labeled `Low-P' is the same object indicated by a green circle in 
Figure~\ref{fig:CFHT4} and is discussed in Appendix~\ref{lowp_star}. 
\label{fig: PE}}
\end{figure}

Figure~\ref{fig: PE} displays PE versus A$_{V_O}$ for the objects in the two
fields observed here, color coded as indicated in the inset legend. The objects appearing
in this plot all met five selection criteria: $\sigma_{EPA} < 28.65\degr$ (i.e., $P^\prime / \sigma_P > 1$);
$\sigma_{(H-M)} < 0.3$~mag (where $m_H$ was from the Mimir observations and
$m_M$ was from WISE); $\sigma_P < 2$\%; (A$_V / \sigma_{A_V}) > 1$ 
\citep[where A$_V = 7.6 $~($H - M - 0.08$)~mag;][]{Majewski11}; and
(PE $/ \sigma_{PE}) > 1$. 
The target BDs and YSOs were excluded from the
power law fits, and the fits were weighted by the propagated $\sigma_{PE}$ variances.
The 2M0444 field shows a fairly steep PE versus A$_{V_O}$ slope of
$-0.81 \pm 0.15$.
The more extincted CFHT~Tau~4 field shows a shallower $-0.54 \pm 0.11$ slope,
indicating a weaker loss of PE with optical depth than seen in the other field,
even though the CFHT~Tau~4 field was probed to higher A$_{V_O}$.
This could indicate different intensities or SEDs of the external illumination
for the two fields, different dust size distributions in the fields, or some combination
of both sets of effects.

Examination of Figure~\ref{fig: PE} leads to a first conclusion that PE in these fields is 
not independent of A$_{V_O}$ (slope is not zero) nor does
$P^\prime_O$ exist only at the cloud surface (slope is not $-1$). A second conclusion
is that the locations of ITG~16 and ITG~19 close to the green line in Figure~\ref{fig: PE} are consistent with
their assignments as normal background objects and not embedded YSOs (see Appendix~\ref{not_YSOs}).
ITG~15, ITG~17, ITG~25, and 2M0444 all exhibit much lower polarization percentages
for their apparent ($H - M$)-based A$_{V_O}$ values, compared to normal background objects.
As much of the apparent reddening for these objects is due to thermal emission from
their warm disks, some portion of their PE departures is due to this effect. 
However, they also show weaker observed polarizations due to 
their intrinsic (disk) polarizations being diminished by passage through the magnetized
HCl2 material. The green circled labeled `Low-P' identifies a star with a lower polarization percentage
($P^\prime_O$ $= 0.69 \pm 0.56$\%) than objects of similar brightness in the CFHT~Tau~4 field.
Its nature is examined in Appendix~\ref{lowp_star}.

The decay of PE with A$_V$ may be interpreted under the RATs paradigm as loss of the radiation
needed to spin up dust grains deeper in the cloud interior, with the assumption that the radiation
arises outside of the cloud. 
If the illuminating radiation arrives to one side of the cloud, only,
then the power laws of Figure~\ref{fig: PE}, with indices $\alpha$, maybe inverted to predict
polarization fractions $P$ that scale like A$_V^{\alpha + 1}$.

For HCl2, there is no obvious illuminator providing such anisotropic radiation, so a more likely model
is one where the Diffuse Galactic Light \citep[e.g.,][]{DGL} or the Interstellar Radiation Field 
\citep[e.g.,][]{ISRF} illuminates the cloud from both front and back sides. 
The model of a slab immersed
in two-sided uniform illumination yields a functional form for $P(A_V)$ that changes strongly at both surfaces and less strongly
in the central portion of the cloud. Under this model, the foreground cloud polarization that
is added to the target intrinsic polarization will depend on depth of the target into the cloud as:
\begin{equation}\label{eq: 1}
P / P_{max} =  0.5 \  \{1 + R_1^{(\alpha + 1)}  - (1 - R_1)^{(\alpha + 1)}\},
\end{equation}
\noindent
where $P / P_{max}$ is the fractional polarization signal contributed, relative to the maximum
along that line of sight, and $R_1$ is the similar ratio of the extinction A$_V$ to the target, relative
to the maximum extinction along the line of sight. 

\begin{figure}
\center
\includegraphics[width=5in,trim= 0.3in 0.55in 0.1in 0.31in]{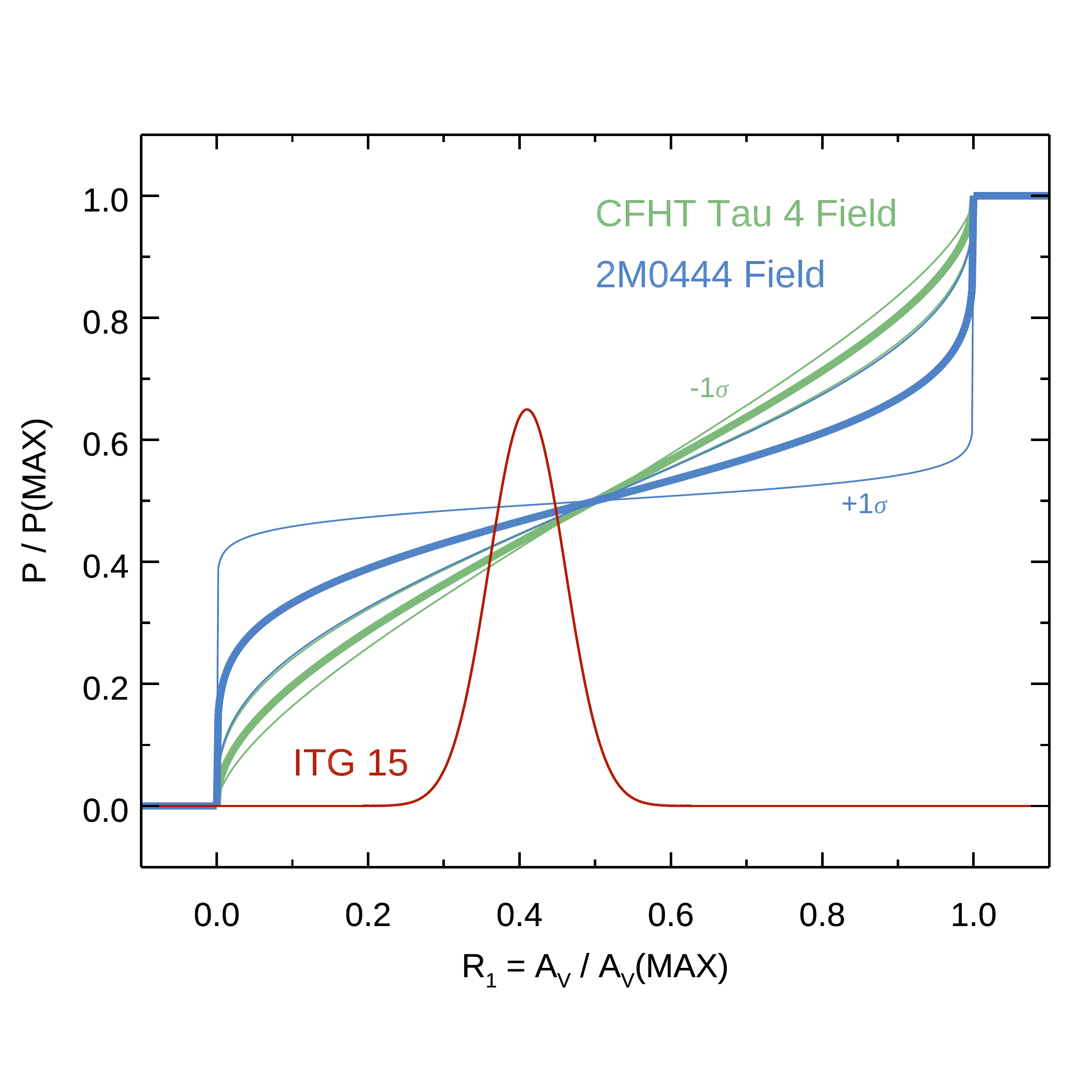}
\caption{Plot of normalized polarization percentage contributed to the starlight of an embedded
target versus normalized visual extinction to the target into the
cloud, $R_1$. Thick green curve shows the behavior of Eq~\ref{eq: 1} for the CFHT~Tau~4 star
field, derived from the power law fit shown in Figure~\ref{fig: PE}. Thinner green curves show
the behavior for slope values ($\alpha$) that are offset from nominal by the uncertainty in the fitted slope.
Blue thick and thin curves show the similar behaviors for the 2M0444 field. Red curve shows a
gaussian representation of the $R_1$ value and uncertainty listed for ITG~15 in 
Table~\ref{tab:Av}. The effective correction factors for the Stokes $U$ and $Q$ values were
found from the integrated overlap of the ITG~15 curve with one of the green curves representing the
PE behavior for the field containing that star, as described in the text.
\label{fig: P_vs_A}}
\end{figure}

Figure~\ref{fig: P_vs_A} presents the Eq~\ref{eq: 1} curves for the $\alpha$ values displayed
in Figure~\ref{fig: PE} for the two observed fields. The thicker green and blue curves show
the behavior for $\alpha = -0.54$ (the CFHT~Tau~4 field) and $\alpha = -0.81$ (the 2M0444 field).
Thinner curves flanking the thick curves show the behavior for $\alpha$ values offset by
$\pm 1\sigma$ from nominal. Values of $R_1$ less than zero and beyond unity are
assumed to produce no foreground cloud polarization and full foreground cloud polarization,
respectively.

% Fig 9: P_vs_A model

To develop the best estimates of the HCl2 contributed foreground polarization for each of the
four target systems, a Monte Carlo simulation was employed. For each target system, it 
started with the appropriate slope for the observed field and added gaussian deviates
scaled by the uncertainty in the slope to produce a trial slope. The A$_V$ likelihood function
for each target was taken as a gaussian with the mean and width as listed for the value
and uncertainty of $R_1$ in Table~\ref{tab:Av} (and as displayed in Figure~\ref{fig: P_vs_A}). 
The $P(A_V)$ curve was integrated with
the A$_V$ likelihood as a kernel to yield one $P / P_{max}$ value. The loop of selecting
another trial slope to yield a new $P / P_{max}$ value was repeated to develop a distribution
function for that ratio, and afterwards characterized as a mean and standard deviation
for the ratio.

The Stokes $U$ and $Q$ parameters
were assumed to follow the behavior of the polarization with extinction, such that the apportioned foreground 
HCl2 contributed Stokes $U_F$ was formed
as the product of the $P / P_{max}$ ratio and the full interpolated background star Stokes $U_I$,
and similarly to create $Q_F$. The similarly scaled interpolated background star Stokes parameter
uncertainties were added in quadrature with the dispersion of the $P / P_{max}$ probability
function to yield final uncertainties for $U_F$ and $Q_F$.

\subsubsection{Applying the Foreground Corrections}\label{subsec:correct}

These apportioned foreground Stokes parameters were subtracted from the observed Stokes
parameters for a particular target system to yield residuals ($Q_R$, $U_R$), and uncertainties 
were propagated. These residual values represent the current best estimates for the intrinsic Stokes
parameters of each of the target YSO and BD systems, without contamination from the
HCl2 contributions.

This foreground correction process is summarized in Table~\ref{tab:UQ}, which lists for each of the objects: the observed 
Stokes parameters ($Q_O$, $U_O$); the Stokes parameters estimated from the
interpolation of the background star values across each of the fields of view at the positions
of each of the BDs and YSOs ($Q_I$, $U_I$); the Stokes parameters apportioned to the foreground
of each object, using the analysis based on Eq~\ref{eq: 1}, ($Q_F$, $U_F$); and the residual differences
of the observed and apportioned foreground sets of values ($U_R$, $Q_R$) that reveal 
the intrinsic NIR polarization of the YSO and BD systems.

% Table 4 Stokes U, Q corrections
\begin{deluxetable}{llcccccc}
\tablecaption{Stokes Parameters: Observed, HCl2 Foregrounds, and Residuals \label{tab:UQ}}
\tablewidth{7.5truein}
\tablehead{\\
\colhead{Desig.}&\colhead{Quantities}&\colhead{Stokes $Q$}&\colhead{Stokes $U$}&\colhead{$P$}&\colhead{$\sigma_P$}&\colhead{$P^\prime$}&\colhead{EPA}\\
&&\colhead{(\%)}&\colhead{(\%)}&\colhead{(\%)}&\colhead{(\%)}&\colhead{(\%)}&\colhead{(\degr)}\\
\colhead{(1)}&\colhead{(2)}&\colhead{(3)}&\colhead{(4)}&\colhead{(5)}&\colhead{(6)}&\colhead{(7)}&\colhead{(8)}
}
\startdata
ITG~15 & A: Observed ($X_O$)$^{\rm a}$&  $-$0.44$\pm$0.08 & $-$0.10$\pm$0.08 & 0.45 & 0.08 & 0.44 & 96.2$\pm$5.4\\
		& B: Interpolated Background ($X_I$) & $-$1.22$\pm$0.24 & $+2.82\pm$0.25 & 3.07& 0.25&3.06&56.7$\pm$2.2\\
            & C: Apportioned Foreground ($X_F$) & $-$0.54$\pm$0.11 &+1.24$\pm$0.11 & 1.35 & 0.11 & 1.35 & 56.8$\pm$2.3\\
          & D: Residuals (Intrinsic = A - C) ($X_R$)&+0.10$\pm$0.14 & $-$1.34$\pm$0.13 & 1.34 & 0.13 & 1.34 &  137.0$\pm$2.8\\
          & E: PA Difference (= D - B) ($X_D$)&&&&&& 80.2$\pm$3.6\\[10pt]
ITG~17 &  A: Observed ($X_O$)&  $-$0.24$\pm$0.11 &+0.16$\pm$0.11 & 0.29 & 0.11 & 0.26  & 73.0$\pm$11.5\\
		& B: Interpolated Background ($X_I$) & $-1.31\pm0.27$ & $+2.83\pm0.26$&3.12&0.26&3.11&57.4$\pm$2.3\\
            & C: Apportioned Foreground ($X_F$)& $-$0.68$\pm$0.14 &+1.47$\pm$0.14 & 1.62 & 0.14 & 1.61 & 57.4$\pm$2.5\\
            & D: Residuals (Intrinsic = A - C)  ($X_R$)&+1.08$\pm$0.29 & $-$2.67$\pm$0.28 & 2.88 &  0.28 & 2.87 & 146.0$\pm$2.7\\
            & E: PA Difference (= D - B) ($X_D$)&                         &                            &         &         &        & 88.6$\pm$3.7\\
            & F: ALMA  &&&&&&25$\pm$5$^{\rm b}$\\
            & &&&&&&40$\pm$10$^{\rm c}$\\[10pt]
ITG~25 & A: Observed ($X_O$)&+2.04$\pm$0.10 &+3.53$\pm$0.10 & 4.08 & 0.10 & 4.08 & 30.0$\pm$0.7\\
		& B: Interpolated Background ($X_I$) & $+1.41\pm0.42$ & $+4.68\pm0.42$ & 4.89 & 0.42 & 4.87 & 36.6$\pm$2.4\\
            & C: Apportioned Foreground ($X_F$)&+1.03$\pm$0.31 &+3.42$\pm$0.34 & 3.57 & 0.33 & 3.55 & 36.6$\pm$2.7\\
            & D: Residuals (Intrinsic = A - C) ($X_R$)&+1.01$\pm$0.33 & 0.11$\pm$0.35 & 1.02 &  0.34 & 0.96 & 3.0$\pm$10.1\\
            & E: PA Difference (= D - B) ($X_D$)&                         &                            &         &         &        & 146.4$\pm$10.5\\[10pt]
2M0444 & A: Observed ($X_O$)&+0.51$\pm$0.23 &+1.21$\pm$0.24 & 1.31 & 0.24 & 1.29 & 33.6$\pm$5.3\\
		& B: Interpolated Background ($X_I$) & $+0.96\pm0.10$ & $+1.80\pm0.10$ & 2.04 & 0.10 & 2.04 & 31.0$\pm$1.4\\
            & C: Apportioned Foreground ($X_F$)&+0.43$\pm$0.05 &+0.81$\pm$0.05 & 0.92 & 0.05 & 0.92 & 31.0$\pm$1.6\\
            & D: Residuals (Intrinsic = A - C) ($X_R$)& +0.08$\pm$0.23 & +0.40$\pm$0.25 & 0.41 & 0.24 & 0.33 & 39.5$\pm$20.8\\
            & E: PA Difference (= D - B) ($X_D$)&                         &                            &         &         &        & 8.5$\pm$20.9\\
            & F: ALMA &&&&&&115$\pm$15$^{\rm b}$\\
            &&&&&&&70$\pm$10$^{\rm c}$\\
\enddata
\tablecomments{References: $^{\rm a}$~Indicator of subscript label for all row quantities, $^{\rm b}$~\citet{Ricci14}, $^{\rm c}$~\citet{Rilinger19}
}
\end{deluxetable}

The process steps for the ITG~15 object began with the first (A) row for that object in 
Table~\ref{tab:UQ} which lists the observed Stokes parameters and the polarization parameters derived 
from them, all with implied ``O'' subscripts as noted. 
Those observed $H$-band Stokes $Q_O$ and $U_O$ were $-0.44\% \pm 0.08$\%
and $-0.10\% \pm 0.08$\%, respectively. 
The second (B) row lists the interpolated values for
Stokes $Q_I$ and $U_I$ at the position of ITG~15 as $-1.22\% \pm 0.24$\% and $+2.82\% \pm 0.25$\%,
respectively.
The third (C) row lists the Eq~\ref{eq: 1} apportionment of the B-row values to yield the HCl2 foreground
corrections of $-0.54\% \pm 0.11\%$ and $+1.24\% \pm 0.11\%$ for $Q_F$ and $U_F$, respectively.
Subtracting these from the observed values produced the (D) row of residuals with
Stokes $Q_R$ and $U_R$ values of $+0.10\% \pm 0.14$\% and $-1.34\% \pm 0.13$\%, respectively.
From each of these four sets of Stokes $Q$ and $U$ values, the derived quantities $P$, 
$\sigma_P$, \Pp, EPA, and $\sigma_{EPA}$ were developed, and are listed in columns (5) through (8) of
the (A)-(D) rows
for each object in Table~\ref{tab:UQ}. Note that once corrected for the apportioned Stokes
foreground $Q_I$ and $U_I$
contributions due to the passage of light through HCl2, the residual debiased polarization
percentage $P^\prime_R$ for ITG~15 
rose to $1.34\% \pm 0.13$\% from the observed $P^\prime_O$ value of
$0.44\% \pm 0.08$\%.

Similarly, the observed polarization EPA$_O$ (in row A) for ITG~15 changed from 
$96\degr.2 \pm 5\degr.4$ to an 
EPA$_R$ of $137\degr.0 \pm 2\degr.8$ 
upon correction for the HCl2 contributions. The interpolated EPA$_I$
(in row B) is a direct measure of the plane of sky magnetic field orientation for HCl2, so it can also
be compared to the residual EPA$_R$ (in row D) to assess any relationship of the intrinsic
polarization of this BD disk system to the ambient magnetic field. That
EPA$_D$ difference (EPA$_R$ - EPA$_I$) is listed in the (E) row for ITG~15 as
$80\degr.2 \pm 3\degr.6$, 
which is close to being perpendicular.

The Table~\ref{tab:UQ} entries for the remaining three BD and YSO objects follow the same
order of processing steps and comparisons to their ambient magnetic field orientations.
For the two BDs, ITG~17 and 2M0444, the EPAs of ALMA-traced dust disk orientations were
determined by \citet{Ricci14} and \citet{Rilinger19} and are listed in the (F) rows.

Changing the angular resolution of the interpolated HCl2 Stokes maps had only minor effects
on the residuals listed in the (D) rows for each object in Table~\ref{tab:UQ}. Such changes tended to be at
or below the 1~$\sigma$ level for EPA$_R$ and much less than 1~$\sigma$ for $P^\prime_R$
values when the interpolation resolution was changed by $\pm$1~arcmin FWHM, for example.
Thus, the residual values are robust against the interpolation resolution
chosen.

\subsection{Intrinsic Near-Infrared Polarization of Brown Dwarf and YSO Disks} \label{subsec:disks}

The debiased polarization percentage residuals for one of the two BDs (ITG~17) and both of the YSOs are all significant at,
or beyond, the 2.8~$\sigma$ level while the other BD (2M0444) has only a $1.4 \sigma$ residual. 
These residuals represent the current best estimate of the 
polarization emitted from these systems prior to that radiation being modified by the
magnetized dust within HCl2. 
Hence, these residuals are the polarization properties of the 
light arising from the photospheres and disks of each of these systems. In three of the four cases,
significant polarization was found, which is highly unlikely to arise in their photospheres and
so must arise from their disks (none of the four systems exhibit envelope emission in their 
SEDs; \citealt{Andrews13}).

\begin{figure}
\includegraphics[width=0.97\textwidth,trim= 0.35in 0.2in 0.2in 0.0in]{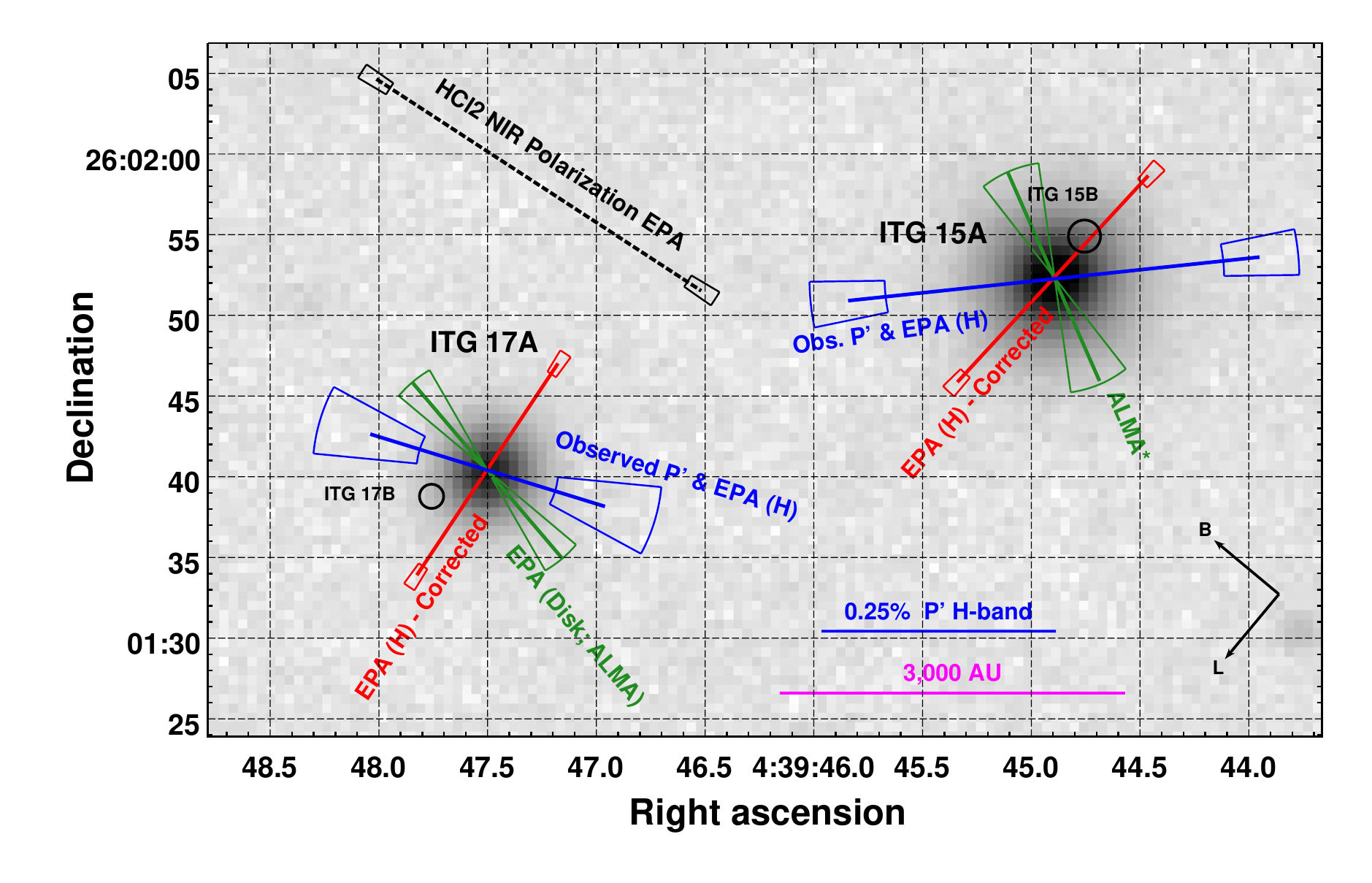}
\caption{Zoom of the central region of Figure~\ref{fig:CFHT4}, showing the relative 
locations of ITG~17A and ITG~15A as well as their possible companions ITG~17B and ITG~15B. 
The $H$-band observed $P^\prime_O$ and 
EPA$_O$ values
are encoded in the lengths and orientations of the blue lines, with a 0.25\% blue reference
line shown at lower right. The EPA of the
orientation of the elongation of the ITG~17 disk modeled using ALMA continuum imaging analysis by
\citet{Rilinger19} is indicated as the green line. The EPA of the orientation of the ITG~15 disk 
(labeled ALMA*) deduced from the observations obtained by \citet{WD18} is also shown in green and is described in Section~\ref{sec:predict} of the text. After correcting the
observed $H$-band Stokes parameters for the foreground HCl2 field polarization, whose average EPA$_I$ orientation
is shown as the black dotted line, the intrinsic EPA$_R$ values for ITG~17 and 
ITG~15 are recovered and are shown as the red lines. Disk rotational angular momenta {\bf J}
would be parallel to these red EPA$_R$ lines and perpendicular to the magnetic field in HCl2, as noted in the text. 
Uncertainties in $P^\prime$ (actual and arbitrarily scaled)
and EPA are indicated by the torus segments at the ends of each line, representing $\pm 1 \sigma$ ranges.
The ALMA line lengths have no meaning, hence their error tori extend to zero length to indicate that
the angular uncertainties do have meaning.
At 140~pc distance, the
projected relative separation of ITG~17A and ITG~15A is about 5,200~au. The projected
relative separations of ITG~15A from ITG~15B and ITG~17B from ITG~17A are about 400-600~au.
\label{fig: CFHT_zoom}}
\end{figure}

Figure~\ref{fig: CFHT_zoom} shows a zoomed portion of the CFHT~Tau~4 field
presented in Figure~\ref{fig:CFHT4}. This portion includes both of the ITG~15AB and ITG~17AB 
systems and shows where other studies located the faint (B) possible companions
(Table~\ref{tab: prop}).
The blue pseudo-vector lines indicate the observed NIR polarization $P^\prime_O$ and EPA$_O$ values for 
the primary (A) objects.
The dashed black line shows the
orientation of the mean interpolated polarization EPA$_I$ for this region, i.e., the magnetic
field orientation in this portion of HCl2.

The EPAs observed for
ITG~17 (EPA$_O$) and for HCl2 (EPA$_I$) differ by only 
$16\degr \pm 12$\degr,
 implying similarity, but their fractional polarizations $P^\prime_O$ and $P^\prime_I$ differ by 
 $3.0\% \pm 0.4$\%,
  which is significant. ITG~15 shows a 
greater EPA deviation (EPA$_O$ - EPA$_I$) from that of HCl2 in this region, namely 
$39\degr \pm 6$\degr\ and
a polarization fraction difference 
($P^\prime_O$ - $P^\prime_I$ = $2.6 \pm 0.3$\%)
nearly as large as is seen for ITG~17. Thus, the {\it observed} polarizations
for these objects do not match the polarization created by magnetically-aligned dust grains in HCl2.
The light from these two disk systems does have to pass through portions of the HCl2
aligned dust grains, as argued earlier. Correcting for this HCl2 contribution results in the 
intrinsic (residual) polarization EPA$_R$ values listed in the (D) rows in Table~\ref{tab:UQ} 
and shown in Figure~\ref{fig: CFHT_zoom} 
as the red lines\footnote{The red lines shown in Figure~\ref{fig: CFHT_zoom} do not encode the intrinsic 
$P^\prime_R$ values,
as they would greatly exceed the lengths of the observed $P^\prime_O$ values shown as the blue pseudo-vectors. 
Instead, the red lines convey full EPA$_R$ meaning but only a scaled sense of $P^\prime_R$.} 
and labels at each of the two objects. 

% Fig 10

The red lines in Figure~\ref{fig: CFHT_zoom} are remarkable for three reasons. First, both
are nearly perpendicular to the HCl2 magnetic field EPA$_I$, as listed
in the (E) rows in Table~\ref{tab:UQ}. As these intrinsic NIR polarization disk EPA$_R$ values are 
expected to be parallel to their sky projected disk
angular momentum vectors ({\bf J}) (see Section~\ref{sec:discussion}), 
the local magnetic field {\bf B} and disk momenta {\bf J}
are apparently perpendicular for both systems. Second, the
red lines are parallel to each other, as listed in the (D) rows in Table~\ref{tab:UQ} ($\Delta {\rm PA} = 
9\degr \pm 4\degr$), indicating that
the {\bf J} vectors for the two disks are aligned (or anti-aligned, as kinematic information is lacking). 
Finally, the EPA of the elongation of the disk for ITG~17, as modeled for
ALMA continuum imaging by \citet[][namely $40\degr \pm 10$\degr]{Rilinger19}, is shown as the green line and label
in Figure~\ref{fig: CFHT_zoom}.
It is $106\degr \pm 10$\degr\ 
from the intrinsic (residual) NIR EPA$_R$ for ITG~17, and so, nearly perpendicular to the NIR EPA$_R$ and thus
perpendicular to the inferred {\bf J} vector. 
Hence, for ITG~17, both ALMA continuum imaging and NIR polarimetry agree as to the disk orientation.
The
difference angle grows to 121\degr, however, if the orientation EPA of the ALMA disk modeled by \citet{Ricci14}
is instead considered. 

% Fig 11
\begin{figure}
\includegraphics[width=0.97\textwidth,trim= 0.35in 0.2in 0.2in 0.0in]{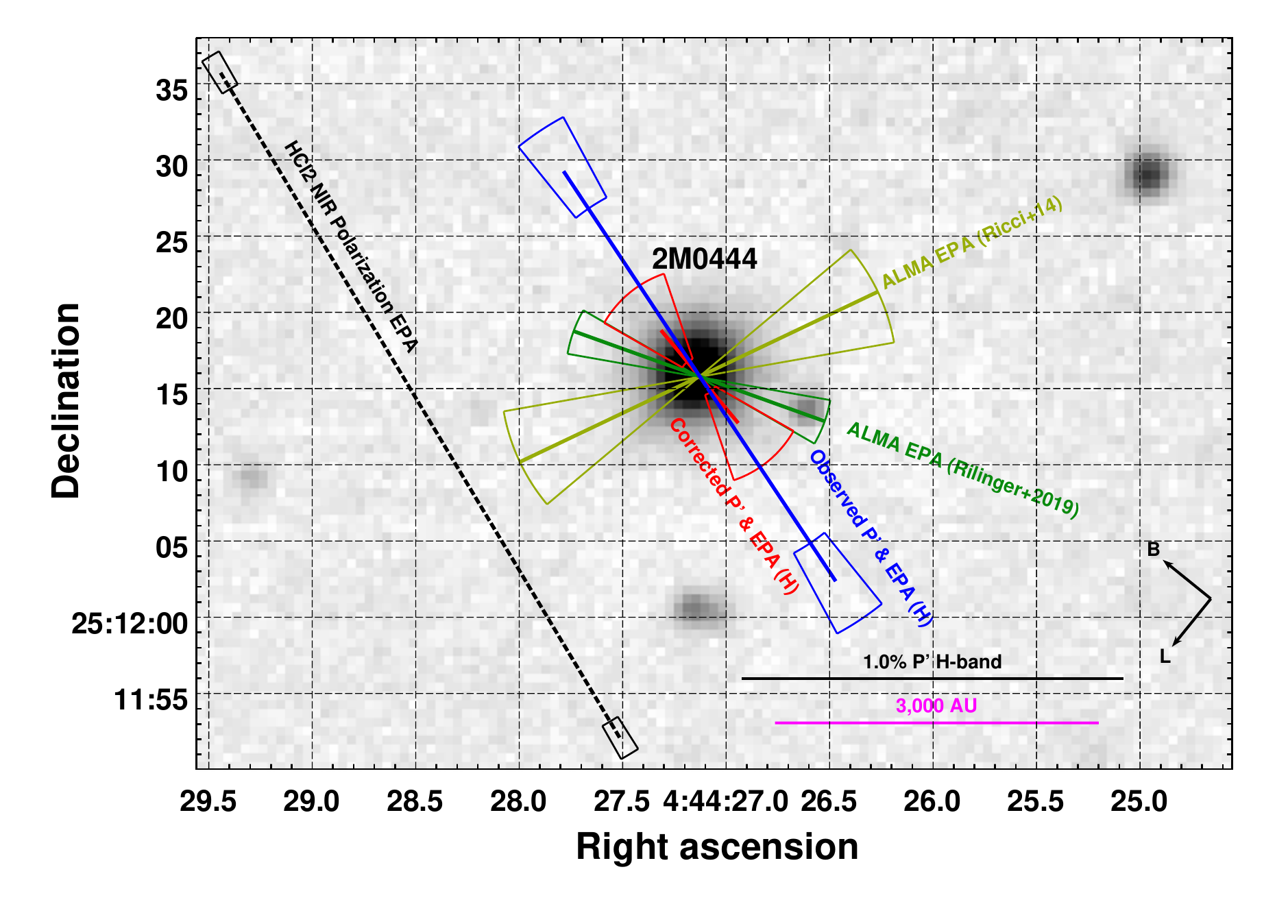}
\caption{Zoom of the central region of Figure~\ref{fig:2M0444} containing the BD 2M0444. 
In this figure, all NIR polarization vectors
are scaled to the black reference at lower right. The NIR polarization direction (EPA$_I$) and 
magnitude $P^\prime_I$ for this direction
through HCl2 are indicated by the dashed black line at left. The observed $H$-band polarization is shown as the
blue line. The 2M0444 disk $H$-band intrinsic polarization is shown as the red line, which should be perpendicular
to the disk elongation direction. Uncertainties in $P^\prime$ 
and EPA are indicated by the torus segments at the ends of each line, representing $\pm 1 \sigma$ ranges.
The ALMA disk orientations on the sky are indicated by the dark green line for the \citet{Rilinger19}
imaging and by the lighter green line for the \citet{Ricci14} CO moment determination. Line lengths representing ALMA
findings have arbitrary scaling: the angular uncertainties do carry meaning. The NIR inferred 
disk orientation (EPA$_R$) is perpendicular to that found by \citet{Ricci14} from their CO moment map.
\label{fig: 2M_zoom}}
\end{figure}

A similar comparison of NIR and ALMA findings for 2M0444 is shown as Figure~\ref{fig: 2M_zoom}.
There, the EPA$_I$ of the HCl2 polarization (black dashed line) is seen to be parallel to the observed $H$-band polarization 
orientation (EPA$_O$; blue line),
but shows a higher fractional polarization ($P^\prime_I$ versus $P^\prime_O$). The Stokes differencing results in the red, corrected polarization
line that encodes $P^\prime_R$ and EPA$_R$ in the Figure. 
The large extents of the red error tori signal the low significance of the NIR residuals, however.
The two determinations of the disk orientation from ALMA observations are shown as the dark green line
for the \citet{Rilinger19} resolved continuum imaging and as the lighter green line for the \citet{Ricci14} 
CO moment map fitting. The weak apparent agreement of the corrected NIR (red) and ALMA continuum (dark green)
orientations actually signals disagreement, as disk elongation 
is interpreted to be perpendicular to NIR intrinsic polarization. The NIR and \citet{Ricci14} findings are
closer to being perpendicular, with an implied disk orientation difference angle of $76 \pm 26$\degr, 
only $0.5 \sigma$ from 90\degr, though with high angular uncertainty. 
The NIR to ALMA continuum difference is $31 \pm 29$\degr, 
some $2 \sigma$ from 90\degr. Both comparisons are likely too weak to support strong conclusions
regarding the disk alignment for 2M0444.

\subsubsection{Implications for Disk Orientations}\label{sec:predict}

No previously published analysis of ALMA observations of the ITG~15 YSO system provide 
modeling sufficient to allow direct 
determination of the dust elongation EPA or gas angular momentum {\bf J} for its disk. 
ALMA observations by \citet{WD18} do show a dust detection map,
but lack deconvolution to establish disk orientation. Detailed SED fitting by \citet{Ballering19} accounted
for inclination but not for disk orientation. 
Based on the correlation of the intrinsic $H$-band EPA$_R$ values for ITG~15 and ITG~17, 
and the near perpendicular nature of the ALMA EPA and NIR EPA$_R$ for ITG~17
under the \citet{Rilinger19} model, 
a prediction for the ITG~15 dust disk elongation EPA of about $47\degr \pm 3$\degr\ can be made. 

The ALMA continuum data for ITG~15 system obtained for the \citet{WD18} study were fetched from the
ALMA archive. Gaussian fitting after deconvolution by the sampling beam yielded major and minor axes
FWHM of $271 \pm 35$~mas and $149 \pm 45$~mas, respectively, at a position angle of $23 \pm 15$\degr\
for ITG~15A. The fainter secondary ITG~15B showed deconvolved FWHMs of $369 \pm 117$~mas
and $105 \pm 85$~mas at a position angle of $13 \pm 24$\degr. This new ALMA disk PA is 
indicted in the Figure~\ref{fig: CFHT_zoom} zoom image by the `ALMA*' label.

The NIR prediction and ALMA deconvolution differ in their estimated plane of sky disk EPA by
$24 \pm 15$\degr\, or about 1.6~$\sigma$. Possibly better data might be ALMA spectral line observations
that are sufficient to establish the orientation and magnitude of the disk angular momentum {\bf J}, as
was done by \citet{Ricci14} for 2M0444 (but not for ITG~15). 

The ALMA-traced disks about ITG~15A and ITG~17A have PAs that differ by $23 \pm 18$\degr, or 
only 1.3~$\sigma$ (compared to the even smaller NIR difference of $9 \pm 4$\degr).
This seems to indicate that both the NIR intrinsic polarizations and the ALMA continuum observations
favor disks about these two systems that are quite similar in sky orientations.
The ITG~15A and ITG~15B disks have PAs that differ by $10 \pm 28$\degr, which is too uncertain for
strong conclusions.

For the ITG~25 system, though the inferred intrinsic NIR
polarization $P^\prime_R$ is weaker than for the ITG~15 and ITG~17 systems, there is sufficient
residual polarization signal to infer the EPA$_R$ orientation of its disk, even though no ALMA observations
have determined the disk elongation EPA and no ALMA observations exist in the archive.
Analogous to the ITG~17 system, the NIR polarization 
predicted elongation EPA for ALMA observations of the ITG~25 disk would be about $93\degr \pm 10$\degr.

\section{Discussion} \label{sec:discussion}

Characterizing the extinctions and polarizations of the many objects background to HCl2, when compared
to the measured values for the ITG systems and 2M0444, led to the conclusion that 
both BDs and both YSOs are located within HCl2. The background star polarizations
were used to establish the Stokes parameters contributions resulting from magnetically aligned dust grains within HCl2.
Removal of these contributions from the measured Stokes parameters for the embedded systems left
residual polarizations that are intrinsic to those systems, most likely resulting from the polarization
produced by scattering of photospheric light from disk surfaces. 

For a net intrinsic polarization to arise from disk systems, three conditions must be present, none
of which are particularly difficult to realize. If the disk is generally symmetric and not dominated by a 
small number of spiral arms, say, then single scattering of photospheric light by disk surface layer dust will produce a 
centrosymmetric polarization pattern, with the prevailing electric field vector perpendicular to the radial
vector from the central object to the scattering location within the disk \citep[e.g.,][]{Silber00,Apai04}. Face-on and nearly face-on
disks show just such a pattern \citep[e.g.,][]{Potter05,Hales06,Follette13}. 
However, for an unresolved disk, a face-on presentation
has net symmetry in the polarization pattern and will result in low to zero net polarization.
Edge-on disks might be expected to extinct scattered light from inner disk surfaces due to high
optical depths in thin disks or due to flared outer disk regions. 
Hence, some inclination of the disk 
is likely needed in order to detect a net polarization for unresolved NIR observations like those reported here. 
Finally, the scattering phase function must favor scattering 
angles near 90\degr\ over lesser and greater angle values: such is expected for Rayleigh scattering. 
Such phase functions have the effect of boosting the polarization
signal from the ansae of inclined disks at the expense of the polarization arising from the front or back 
portions of the disk, though these regions are located closer in projection to the central object than are the ansae. 
Phase functions favoring 90\degr\ scattering have been measured for resolved disks, 
for example in the nearly edge-on HD~35841 debris disk by \citet{Esposito18}, and show maximum polarization fractions of around 
30\%. 

The combination of some disk inclination, centrosymmetric single scattering from disk surfaces, and
phase functions favoring 90\degr\ results in net intrinsic NIR polarization for the unresolved disk systems
studied here. Further, the relatively high intrinsic polarization fractions for ITG~15 and ITG~17, of
about 1-3\%, implies that a significant fraction of the photospheric light is intercepted by their disks. 
The 30\% polarization fraction measured by \citet{Esposito18} in their resolved observations was
computed relative to the total light reflected from the same regions, which was a factor of 10-30 times
less than that emitted by the HD~35841 central object. That is, if the HD~35841 polarization observations were
not spatially resolved, the net system polarization would be more like 1-3\%, similar to the values measured here
for ITG~15 and ITG~17.

The NIR polarization position angle emergent from an unresolved disk system will be dominated by the 
ansae reflection polarization favored by the scattering phase function, leading to {\it measured EPA$_R$ values being
perpendicular to the elongation axis of the inclined disk} and parallel to the sky-projected disk
angular momentum vector {\bf J}. Hence, for the ITG~17 BD system, we expect
the intrinsic NIR polarization EPA$_R$ ($146\degr.0 \pm 2\degr.7$) to be perpendicular to the 
elongation EPA found by \citet{Rilinger19} ($40\degr \pm 10\degr$) from their ALMA analyses and modeling. 
In Figure~\ref{fig: CFHT_zoom}, the two are very nearly perpendicular ($106\degr \pm 10\degr$),
with most of the difference uncertainty arising from the ALMA modeling. Hence, the NIR and
ALMA modeling agree as to the disk elongation angle for the BD ITG~17.

The similar comparison for the 2M0444 BD, both between the two ALMA studies and between
the ALMA studies and the NIR value found here, was significantly less conclusive due
to the weak SNR of the residual NIR polarization, as noted in the previous Section.

Polarization in the NIR could instead arise from shadowing due to warped
inner disks causing anisotropic disk illumination \citep[e.g.,][]{Benisty18}. Time-dependent
polarization behavior could reveal systems with these properties. A cursory test of
the data obtained for this study did not reveal EPA$_O$ changes with time (see Appendix~\ref{time_EPA}).

\subsection{Aligned Disks in a Wide Binary}

The YSO ITG~15 and BD ITG~17 were identified as a possible wide binary system by
\citet[][as their couple number 28]{Joncour17} and the nature of object clustering within 
Taurus has been studied by \citet{Joncour18}.
Both objects are likely members of the \citet{Galli19} HCl2 cluster 14 group of about ten objects that
spans about 7 cubic parsecs, but these two are the closest pair of objects among all of that cluster, in projection. 
They are at the same distance (see Figure~\ref{fig: P_PM}), and if they are in a circular orbit 
residing purely in the plane of the sky, their orbital
period, based on the masses from Table~\ref{tab: prop}, would be about 0.7~million years, similar to
their inferred ages. Such an orbit has a predicted relative tangential velocity
(0.23~km~s$^{-1}$) which is similar to the measured relative tangential velocity 
($0.45 \pm 0.15$~km~s$^{-1}$) from the 
Gaia~EDR3 reported proper motions of the objects. The projected
separation of 5,200~au for the pair is near the limit for M-dwarf wide binaries \citep{Law10}, but is 
contained within the central portion of the distribution of ultra-wide pairs for Taurus determined by \citet{Joncour17}.
The conclusion is that the possible binaries ITG~15AB and ITG~17AB themselves form a (hierarchical) wide binary 
(or quadruple system). Additionally, the primaries
are somewhat more tightly bound to each other than to the remainder of the cluster 14 group
of \citet{Galli19}. 

ITG~15B was detected by Gaia in EDR3
(see Figure~\ref{fig: P_PM}) and in the ALMA observations of \citet{WD18} but ITG~17B has 
not been reported as detected in ALMA dust continuum observations. As the separations of these 
companions from their primaries
are about 400-600~au, disks would be expected around these systems, based on extrapolation of the statistics for
ALMA disks about Class~II stars more massive than M6 in Taurus \citep{Akeson19} to lower mass objects. 
If ITG~17B is significantly less massive than its primary and the secondary disk mass scales like the
secondary star mass, the ITG~17B disk may be too faint for routine ALMA detection levels.
The \citet{Akeson19} study did not recognize ITG~15~--~ITG~17 as a binary in their Class~II study of
singles, binaries, and multiples, though their summary of the ALMA properties of ITG~15 and ITG~17
in their Table~4, when interpreted as a binary,
do exhibit disk masses, disk mask ratios, and disk-to-star mass ratios similar to the other 
wide binaries (1 - 30~arcsec separation) in their sample.

The parallel EPA$_R$ values for the derived intrinsic $H$-band
polarization from these two objects, with a difference EPA of $9\degr \pm 4\degr$ 
as shown in Figure~\ref{fig: CFHT_zoom}, and the ALMA continuum disk orientation EPA agreements
are strong evidence for aligned disks for this hierarchical binary system. 

\subsection{Brown Dwarf and Disk Formation}

Disk and protostar formation appear in modeling studies to be governed by
numerous effects and quantities, including ambipolar diffusion, the Hall Effect,
Ohmic dissipation, angular momentum and mass transport, infall and outflow,
local chemistry, and ionization fractions and rates, as well as grain sizes and
distributions \citep[see review by ][]{Zhao20}. None of these quantities is
directly available for determination in this study. But, the current disk orientations
for ITG~15 and ITG~17, in relation to each other and in relation to the current
local mean magnetic field sky projected orientation for HCl2 at or near the location
of this binary have been determined. How the apparent orthogonality of
{\bf B} and inferred sky projected {\bf J} for these disks relates to, constrains,
or fails to constrain models of protostellar disk formation and evolution are 
difficult to discern from this work alone.

A key question is the degree to which present day disk orientations and
present day magnetic field orientations can inform physical conditions at the
time of formation of the protostars and their disks. For example, although the 
large-scale magnetic field within HCl2 is unlikely to have changed greatly in orientation
and strength over the last 1-2~Myr, magnetic fields in dense cores and 
protostellar envelopes could be significantly different from the fields
in the more diffuse cloud material surrounding the cores. However, the 
study of solar-mass Class~0 objects and their environs by \citet{Galametz18}
found that ``...an ordered B morphology from the cloud to the envelope is
observed for most of our objects.'' This conclusion of unchanged magnetic field orientations
was also reached for the 
more extended environment from 
6,000~au to pc-scales about the low-mass 
Class~0 protostar GF9-2 by \citet{Clemens18}. Yet a recent study of magnetized
filaments in NGC~1333 by \citet{Doi20} instead finds changes in magnetic field 
morphologies inside of 1~pc but (complex) morphological continuity between 
1~pc and 1000~au. In addition to B-field changes,
the disks surrounding ITG~15 and ITG~17 could have changed their projected 
orientations in their lifetimes. Though for them to appear at the present time in a
parallel configuration, after experiencing orientation evolution likely tied to
their host star masses, seems unlikely.

Progress may be found in asking which characteristics of the ITG~15~--~ITG~17
system could be more easily explained if the current misalignment of the
magnetic field and the disks does indicate initial formation conditions.
An emerging consensus \citep[review by][and references therein]{Zhao20} is that magnetic fields aligned with
disk rotation axes may produce conditions ripe for magnetic braking, leading to
weaker disk angular momenta and smaller disk sizes, while misaligned fields may do
the opposite, producing disks with stronger angular momenta and larger
sizes \citep[e.g., Figure~3 of ][]{Galametz20}, though a recent study
in Orion \citep{Yen21} finds no disk size correlation with magnetic field 
misalignment.  
Turbulence may lead to substantial disks, even for the aligned field case
\citep{Gray18}, indicating that magnetic alignment effects alone may not
be sufficient to predict disk or protostar outcomes.
Disks with strong angular momenta
are also prone to fragmentation \citep[e.g.,][]{Wurster19a}, leading to binaries or multiple star 
systems \citep{Zhao18, Rosen19},
many of which retain at least a significant fraction of their individual disk 
masses \citep{Maury19}.

In the case of ITG~15~--~ITG~17, the binary and aligned disk natures
could both be the result of formation in a misaligned magnetic field condition.
Indeed, the very wide separation could have had the effect of retaining
more disk material, relative to the mass of each host star, than if the stars
had less separation. This seems to be born out in the ratio of the secondary (ITG~17A) 
disk mass to secondary star mass versus the ratio of primary (ITG~15) disk
mass to primary star mass, as shown in Figure~12 of \citet{Akeson19}.
The ITG~15~--~ITG~17 binary exhibits the highest such secondary ratio
of all of their binary systems, even for its relatively high primary ratio.
The secondary disk is the one about the BD ITG~17, 
found to be a somewhat large 80~au \citep{Rilinger19}, which could be
explained by a misaligned magnetic field that was present at proto brown dwarf
formation and remains so to this day.

\subsection{Impacts on Previous Studies}

Near-infrared polarimetry has been used in the past to study YSOs and their environs, including in Taurus.
For example, \citet{Tamura89} examined 39 T~Tauri stars for $K$-band (2.2~$\mu$m) polarization, 
using an aperture polarimeter. They detected linear polarization with a median value of about 0.6\%
but with polarization position angles that sometimes showed large differences compared to optical polarizations
and showed a moderate preference for alignment parallel to local magnetic field orientations. However,
\citet{Tamura89} did not perform foreground polarization corrections to their observations. 
\citet{Tamura89} argued against the Taurus molecular clouds as the {\it origin} of the polarizations they measured, based on mean extinctions and a ratio of $K$-band polarization to extinction, but they did not consider that the intervening material could affect how to use the observed EPA and $P$ values to interpret intrinsic source properties. 
If the mean $H$-band cloud polarization percentages of about 3-4\% (see Table~\ref{tab:UQ}) are scaled to $K$-band using an average Serkowski law \citep[e.g.,][]{Serkowski75}, which characterizes the wavelength dependence of polarization, a predicted contribution from the Taurus material of 1-2\% in $K$-band is obtained.
This exceeds the median $P$ value measured by \citet{Tamura89}. Hence, a reanalysis of the \citet{Tamura89} data to establish the {\it intrinsic} polarization for each source would likely find the same sort (and degree) of polarization position angle changes seen here in the $H$-band for the embedded BD and YSO targets. This calls into question the \citet{Tamura89} conclusions, and those of other similar studies, that are based on {\it observed} EPA or $P$ values but lack corrections for intervening magnetized cloud effects. \citet{Tamura89} could not perform these corrections due to the lack of wide-field imaging polarimetric data, hence their findings regarding alignments of disks and outflows are weakened. The novel aspect 
brought about by the new Mimir observations is the ability to accurately characterize the polarization contributions of the intervening Taurus material to the intrinsic polarization of embedded sources and so allow correction from observed values to the intrinsic ones. 

\section{Summary} \label{sec:summary}

Wide-field near-infrared $H$-band imaging polarimetry observations using Mimir were combined with archival photometry and
Gaia~EDR3 distance and proper motion information to ascertain the presence of intrinsic linear polarization
from two brown dwarf disk systems in Taurus that had previously been analyzed and modeled using 
archival ALMA data 
by \citet{Rilinger19} and two YSO disk systems, all of which are in the direction of the 
Heiles Cloud 2 (HCl2) dark molecular cloud. 
Combining the Gaia information with infrared $(H - M)$
colors enabled classifying most of the 400 objects measured for $H$-band polarization as being foreground, 
embedded, or background to the HCl2 dark molecular cloud. All four target objects were found to 
be embedded within HCl2 and had significant $H$-band polarizations detected. Background objects were used to 
ascertain the polarization signals impressed on starlight by HCl2 due to its magnetically aligned dust grains.
Correcting the apparent $H$-band polarization values by the HCl2 polarization contributions revealed the
intrinsic polarizations of one of the BDs (ITG~17) and both YSO systems, while the remaining BD (2M0444)
had low polarization significance after correction.

The NIR polarization-inferred elongation orientation (EPA$_R$) for the disk around the BD ITG~17, $56\degr \pm 3\degr$, is similar to the 
ALMA elongation orientation, $40\degr \pm 10\degr$, found by \citet{Rilinger19}. 
For the BD 2M0444, 
the corrected NIR polarization yields only a weak comparison with ALMA modeled orientations.

The YSO ITG~15 and the BD ITG~17 likely
form a 5,200~au separation M-dwarf wide binary. Both objects show intrinsic disk polarization position 
angles that are parallel to each other, yet are nearly
perpendicular to the local magnetic field orientation within HCl2. 
This configuration could have arisen from misalignment of an initial magnetic field and the disks at the
time of protostar formation. Such misalignment could have also resulted in somewhat larger disk sizes,
as is observed for the BD ITG~17.

Using the multiple background stars in the wide Mimir fields to ascertain the polarization induced by 
HCl2 enabled applying corrections to the brown dwarf and YSO disk system polarizations that led to completely different
findings for their intrinsic polarization, and thereby disk, properties as compared to their observed
NIR ones. Previous studies that ignored these vital corrections may need to be revisited
and their findings reconsidered.

\vspace{-6mm}

\acknowledgments

This research used the VizieR catalog access tool, CDS,
 Strasbourg, France (DOI : 10.26093/cds/vizier), described
in \citet{Ochsenbein00} and
data from the European Space Agency (ESA) mission
{\it Gaia} (\url{https://www.cosmos.esa.int/gaia}), processed by the {\it Gaia}
Data Processing and Analysis Consortium (DPAC,
\url{https://www.cosmos.esa.int/web/gaia/dpac/consortium}). Funding for the DPAC
has been provided by national institutions, in particular the institutions
participating in the {\it Gaia} Multilateral Agreement.
This study used data products from the Two Micron All Sky Survey, 
which is a joint project of the University of Massachusetts and the Infrared 
Processing and Analysis Center/California Institute of Technology, funded by 
NASA and NSF, and data products from the {\it Wide-Field
Infrared Survey Explorer}, which is a joint project of the University of California, Los Angeles,
and the Jet Propulsion Laboratory (JPL)/California Institute of Technology (CalTech) and funded
by NASA. Based in part on data obtained as part of the UKIRT Infrared Deep Sky Survey (UKIDSS).
This paper makes use of the following ALMA data: ADS/JAO.ALMA\#2012.1.00743.S.  
ALMA is a partnership of ESO (representing its member states), NSF (USA) and NINS (Japan), 
together with NRC (Canada), MOST and ASIAA (Taiwan), and KASI (Republic of Korea), 
in cooperation with the Republic of Chile. The Joint ALMA Observatory is operated by 
ESO, AUI/NRAO and NAOJ. The National Radio Astronomy Observatory is a facility of the National Science Foundation operated under cooperative agreement by Associated Universities, Inc.
Comments and suggestions by the five anonymous reviewers led to numerous
improvements.
This study was based on observations using the 1.8m Perkins Telescope Observatory (PTO) in Arizona, owned and operated by Boston University and managed on-site by Dr. M. Hart. Data were obtained  
using the Mimir instrument, jointly developed at Boston University and Lowell 
Observatory and supported by NASA, NSF, and the W.M. Keck Foundation.
This study was supported by grants  
AST~18-14531 (PI: Clemens) and AST~20-09842 (PI: Pillai) from NSF/MPS to Boston University.

\facilities{Perkins Telescope Observatory (Mimir), and CDS (Vizier \citep{Ochsenbein00})}

\software{SAOImage DS9 \citep{2003ASPC..295..489J}; and
 		topcat \citep{2005ASPC..347...29T}
          }

\vspace{7mm}
\clearpage
\appendix

\vspace{-14mm}

\section{The Distant, Non-YSO Natures of Objects ITG 16, 19, and 21}\label{not_YSOs}

The ITG objects 16, 19, and 21 in the CFHT~Tau~4 field were found to be different in nature
than the BD ITG~17 and the two YSOs, ITG~15 and ITG~25, in that field. 
This Appendix summarizes the findings that these three ITG objects
are normal, diskless stars located far beyond HCl2 and are not associated with 
the embedded young objects in that dense molecular cloud.

There are no parallax or proper motion values in Gaia~EDR3 for ITG~19 or ITG~21, making their direct 
distance determinations impossible.  
ITG~16 is in Gaia~EDR3 and exhibits different values for
parallax and proper motion than for the BD and two YSOs in the CFHT~Tau~4 field, as shown in Figure~\ref{fig: P_PM}.

\begin{figure}
\includegraphics[width=\textwidth,trim= 0in 1.0in 0in 0in,clip]{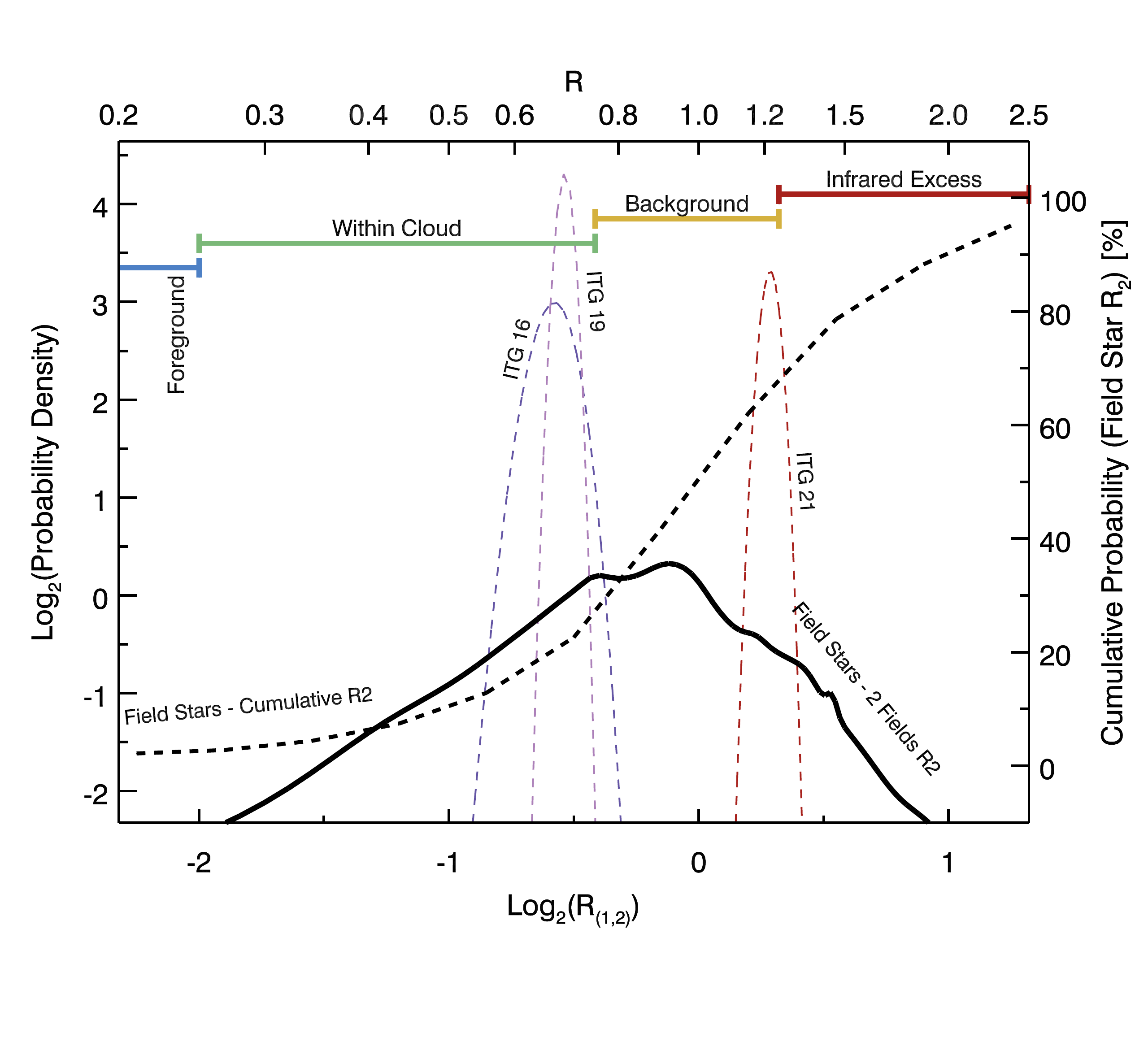}
\caption{Probability distribution functions for the A$_V$ ratio function $R_2$ ($\equiv$ A$_{V_O}$ / A$_{V_I}$), 
similar to the plot shown
as Figure~\ref{fig: R1R2}. 
The solid black curve displays the $R_2$ probability density for the background objects in the two observed
fields. The dashed black curve is the corresponding cumulative $R_2$
probability. Dashed colored curves display the $R_2$ probability densities for the ITG objects 16, 19, and 21. 
Based on these curves, and the Gaia~EDR3 distance to ITG~16, all three are
judged to be more distant than HCl2.
\label{fig: nonYSO_R1R2}}
\end{figure}

All three of these ITG objects lack detailed SED modeling and so have no $(H - M)$ color apportionments 
between disk-based and foreground extinctions.
This leaves only the $R_2$ ratio method (see Section~\ref{subsec:BDs}) for locating the objects relative to HCl2.
Table~\ref{tab:nonYSO_Av} lists, for the three ITG objects, the same information as was provided for
the BDs and YSOs in Table~\ref{tab:Av}, but absent entries for modeled A$_{V_M}$ or $R_1$.
The $R_2$ values were used to place the corresponding gaussian likelihoods for the objects in
Figure~\ref{fig: nonYSO_R1R2}. It shows the
$R_2$ distributions for ITG~16 and ITG~19
weakly favor them being within HCl2. 

However, as already noted in Section~\ref{subsec:dist} and
shown in Figure~\ref{fig: P_PM}, ITG~16 has a Gaia~EDR3 
parallax value placing it about 1~kpc away, versus the 140~pc distance to HCl2. 
ITG~16 does not show the extreme IR excesses
of the three brighter ITG objects (see Figure~\ref{fig: SED}), and shows polarization properties similar to its other sky neighbors (see Figure~\ref{fig:CFHT4}) which are located
at distances well beyond HCl2.
Correcting by the apparent reddening and the Gaia parallax, ITG~16 is judged to be a normal red giant and
not associated with HCl2. 

Although ITG~19 lacks Gaia~EDR3 parallax or proper motion values, it is 
fainter than ITG~16 in the $H$-band and its $R_2$ value 
and its $H$-band polarization properties are very similar to those of ITG~16. These indicate that ITG~19
is also most likely located beyond HCl2.

% Table 3 Av values
\begin{deluxetable}{lcccc}
\tablecaption{Dust Extinctions for the Non-YSO Stars\label{tab:nonYSO_Av}}
\tablewidth{7.5truein}
\tablehead{\\
\colhead{Desig.}&\colhead{A$_{V_I}$}&\colhead{A$_{V_O}$}&\colhead{$^{13}$CO Integrated}&\colhead{$R_2$}\\
&\colhead{Interpolated}&\colhead{Observed}&\colhead{Intensity}&
\colhead{((3)/(2))}\\
&\colhead{(mag)}&\colhead{(mag)}&\colhead{(K~km~s$^{-1}$)}&\\
\colhead{(1)}&\colhead{(2)}&\colhead{(3)}&\colhead{(4)}&\colhead{(5)}
}
\startdata
ITG~16 &  8.6  &  5.8   & 3.3  &  0.67$\pm$0.05   \\[3pt]
ITG~19 & 17.6  & 12.1 & 3.8  &  0.69$\pm$0.02\\[3pt]
ITG~21AB &  9.6  &  11.7 & 3.6 &  1.22$\pm$0.04  \\[3pt]
\enddata
\tablecomments{Uncertainties are about 0.1~mag for column (2), 0.4~mag for column (3), and 0.1~K~km~s$^{-1}$ for
column (4).}
\end{deluxetable}

ITG~21 is a UKIDSS-resolved equal-brightness double (Table~\ref{tab: prop}) but lacks
Gaia~EDR3 parallax and proper motion entries. 
Its $R_2$ value of 1.22 in Table~\ref{tab:nonYSO_Av} falls weakly into the IR excess region, but its broadband 
SED (see Figure~\ref{fig: SED}) reveals that a normal, reddened photosphere dominates, with little apparent 
mid-infrared excess \citep{Bulger14}.
Given its measured $(H - M)$ reddening, if this 
system is also at or beyond 1~kpc distance, then the Table~\ref{tab: prop} spectral type(s) would
be in error and instead should be reassigned to something more like an A main sequence or a late-G giant.

Hence, for ITG~16, ITG~19, and ITG~21, locations within or associated with HCl2 are disfavored.
Instead, all three systems are most likely background to HCl2 and thereby provide representative
magnetic field orientation information for HCl2 in their $H$-band polarization values.

All of the $H$-band polarization objects except the BD and two YSOs in the CFHT~Tau~4 
field have Gaia EDR3 distances placing them beyond the distance to HCl2 (see Figure~\ref{fig: P_PM}).

% Fig 12

\section{Lack of Time-Dependent EPA$_O$ Changes}\label{time_EPA}

An alternative origin for the generation of net NIR polarizations for BD/YSO disk systems 
could involve shadowing due to warped
inner disks causing anisotropic disk illumination \citep[e.g.,][]{Benisty18}. A potential test of the 
application of this scenario to the objects in this study would be detection of time-dependent
polarization properties. Indeed, both ITG~17 (at 2.95~days) and 2M0444 (at 4.43~days) exhibit
periodic optical brightness amplitude variations, as detected using K2 \citep{Howell14} by \citet{Rebull20},
who attributed these to rotation. 

The NIR polarization data obtained here for the six CFHT~Tau~4 field observations (over 3 epochs)
and the four 2M0444 field observations (2 epochs) were examined to test for significant polarization
EPA$_O$ variations. EPA$_O$ deviations from the means, scaled by their uncertainties, were averaged to form a $\chi^2$
quantity for each of the objects in Table~\ref{tab: prop} plus three more comparison objects  
in the 2M0444 field. These comparison objects were chosen to have $H$-band brightnesses similar 
to that for 2M0444.
All objects, except ITG~25, exhibited mean deviations in a narrow range of 0.8--1.2~$\sigma_{EPA_O}$
(i.e., reduced $\chi^2$ values of 0.6--1.5) with no evidence of higher deviation noise for 
ITG~15, ITG~17, or 2M0444 above the values shown
by the other, non-disk ITG and comparison objects. ITG~25 showed a mean EPA$_O$ deviation of 
2.3~$\sigma_{EPA_O}$
(reduced $\chi^2$ of 5.3) but it also had the lowest $\sigma_{EPA_O}$ values of about 1.$\degr$4, versus
the 4--18$\degr$ seen for the fainter objects.  
Although this study was not designed to probe the time-dependent behavior of the NIR 
polarization from BD or YSO disks, to the limits of these current observations, strong deviations
in the polarization EPA$_O$ values were not found.

\section{The Low-Polarization Object J04395: SED Comparisons}\label{lowp_star}

One star in the CFHT~Tau~4 field, 2MASS J04395361+2557485 (hereafter J04395; number 10094 
in Table~\ref{tab:data}), showed low polarization percentage {\bf ($P^\prime_O < 1$\%)} and
moderately high extinction (A$_{V_O} \sim 9$~mag) based on its apparent $(H - M)$ color.
The resulting polarization efficiency is nearly as low as the values seen for the ITG~15 YSO and
ITG~17 BD, as noted in Section~\ref{subsec:PE} and as seen in Figure~\ref{fig: PE}. 
For the BD and YSO, their low PEs are the result of both intrinsic polarizations being
modified by the HCl2 magnetic field polarization and the mid-infrared thermal emission
contributions from the disks of the objects. Could J04395 also have a disk? Is J04395 also embedded in 
HCl2?

The apparent polarization properties of the object were corrected using the bulk Stokes correction method described
in Section~\ref{subsec:methods} and the results are listed in Table~\ref{tab:lowp_UQ}. The derived EPA$_D$ of 
$82.9 \pm 16.9$\degr\ for J04395 suggests that the
disk polarization orientation may be perpendicular to the HCl2 magnetic field
orientation. If it is embedded within HCl2, then this finding matches those for
the other ITG objects. However, Gaia EDR3 lists no parallax or proper motion for this object,
so its location remains unknown. No other objects in the CFHT~Tau~4 or 2M0444 fields
show PE values as low as seen for J04395. 

% Table 5 Stokes U, Q corrections
\begin{deluxetable}{llcccccc}
\tablecaption{Stokes Parameters for J04395 \label{tab:lowp_UQ}}
\tablewidth{7.5truein}
\tablehead{\\
\colhead{Desig.}&\colhead{Quantities}&\colhead{Stokes $Q$}&\colhead{Stokes $U$}&\colhead{$P$}&\colhead{$\sigma_P$}&\colhead{$P^\prime$}&\colhead{EPA}\\
&&\colhead{(\%)}&\colhead{(\%)}&\colhead{(\%)}&\colhead{(\%)}&\colhead{(\%)}&\colhead{(\degr)}\\
\colhead{(1)}&\colhead{(2)}&\colhead{(3)}&\colhead{(4)}&\colhead{(5)}&\colhead{(6)}&\colhead{(7)}&\colhead{(8)}
}
\startdata
J04395 & A: Observed {\bf ($X_O$)} &$-$0.56$\pm$0.56 &+0.70$\pm$0.57 & 0.90 & 0.56 & 0.69 & 64.5$\pm$23.3\\
            & B: Interpolated {\bf ($X_I$)}&$-$0.68$\pm$0.20 &+1.86$\pm$0.20 & 1.98 & 0.20 & 1.97 & 55.1$\pm$2.8\\
            & C: Residuals (Intrinsic) {\bf ($X_R$)}& $+$0.12$\pm$0.59 & $-$1.16$\pm$0.61 & 1.17 & 0.60 & 1.00 & 138.0$\pm$16.7\\
            & D: Difference (= C - B) {\bf ($X_D$)}&                         &                            &         &         &        & 82.9$\pm$16.9\\
\enddata
\end{deluxetable}

\subsection{Spectral Energy Distributions}\label{subsec:SEDs}
\begin{figure}
\includegraphics[width=\textwidth,trim= 0.1in 1.10in 0.1in 0.0in]{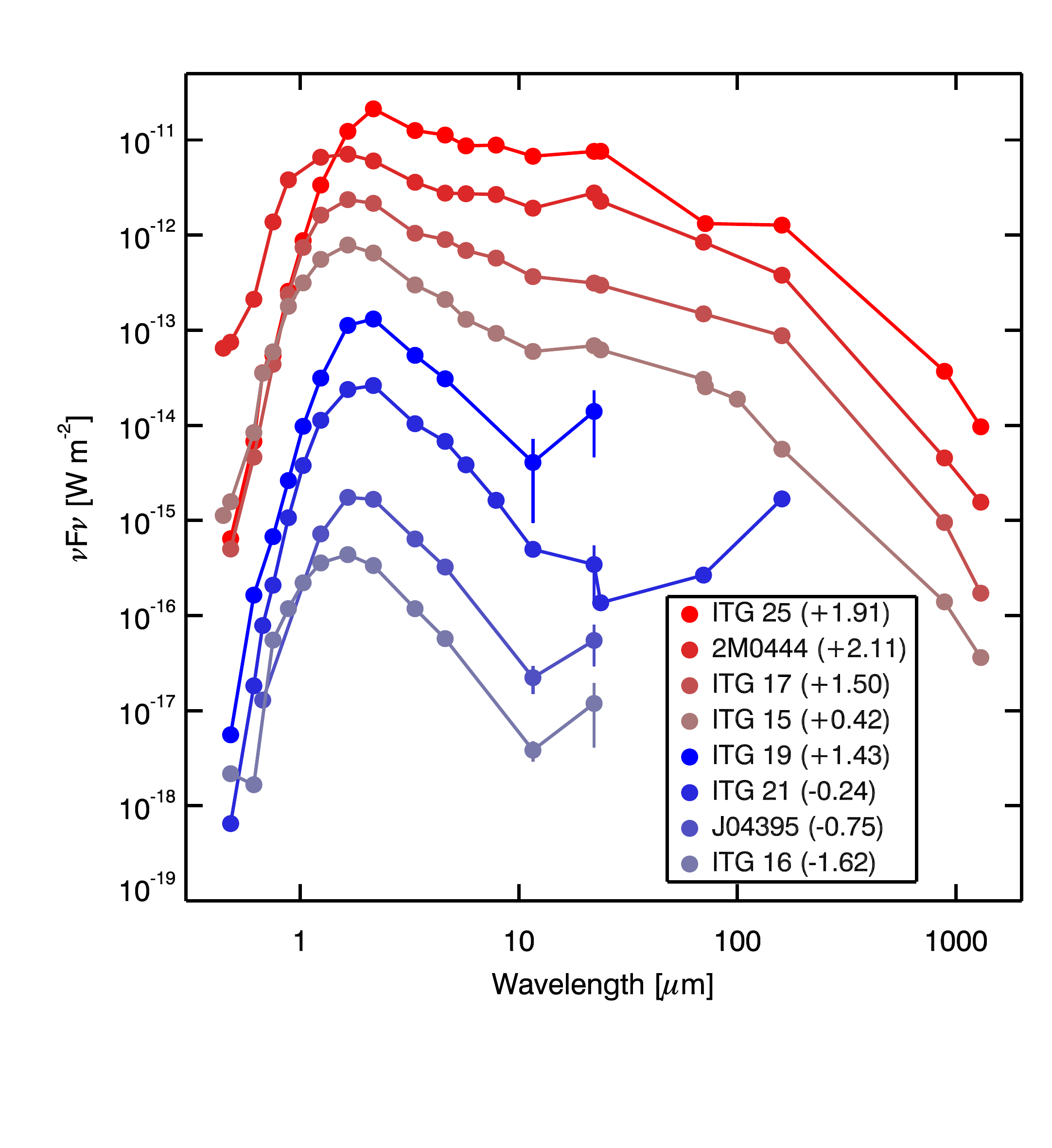}
\caption{Spectral energy distributions (SEDs) of the two BDs, two YSOs, three non-YSO ITG stars, and the 
low-polarization star J04395. The upper four SEDs for the BDs and YSOs are colored in red 
shades and are ranked and offset to
highlight long wavelength disk emission. The lower four SEDs for the non-YSOs are colored in blue shades 
and are similarly
ranked and offset, as described in the text. The legend lists the objects and the base-10
power used to scale each SED. 
\label{fig: SED}}
\end{figure}

To try to understand the nature of J04395, SEDs for the BDs, the YSOs, the non-YSO ITG objects, and for J04395 
were developed, using Vizier \citep{Ochsenbein00} to collect published spectral fluxes. The results are
shown in Figure~\ref{fig: SED}. The SEDs for the eight objects were grouped into two sets:
those showing strong disk emission in mid-infrared (MIR) through mm wavelengths 
(ITG~15, ITG~17, ITG~25, and 2M0444) and those
with weak or questionable MIR emission (ITG~16, ITG~19, ITG~21, and J04395).
In Figure~\ref{fig: SED}, the SEDs showing disk emission are colored different shades of
red, while those lacking strong disk emission are colored different shades of blue.
The SEDs were ordered by the strength of the disk MIR emission relative to the
peak photospheric emission, from the strongest (ITG~25) to the weakest (ITG~16).
Offsets were applied to each SED to reduce overlaps in the MIR range. The 
offsets were multiplicative and are listed as the base-10 log power in parentheses
after the identifier for each object in the inset legend in the Figure. The SEDs have not been 
corrected for extinction effects.

The SED for J04395 is generically very similar to the other three in its diskless or
weak-disk set of SEDs. There is some hint of an upturn at 22~$\mu$m in the
WISE W4 band, which might imply a colder disk or envelope component, 
but the signal-to-noise is less than two, so evidence for any disk is weak. Longer wavelength
data are not available and the star is relatively faint, so confidently resolving whether a 
disk is present will require some effort. 

% Fig 13

The lack of strong disk emission
in the MIR means that the observed $(H - M)$ reddening is fully due to dust 
extinction through HCl2. The A$_V$ map for the CFHT~Tau~4 field of Figure~\ref{fig: CFHT_Av}
predicts a total A$_{V_I}$ through HCl2 at the position of J04395 of 9.8~mag, which 
gives a ratio $R_2$ of observed to interpolated A$_V$ of 0.90. This falls in the
`Background' region designation shown in Figure~\ref{fig: R1R2}. 
Thus, J04395 does not appear to be embedded within HCl2 and
instead is behind the cloud. Hence, J04395 is not an embedded YSO or BD within
the cloud and does not appear to exhibit emission from a strong disk or dusty envelope.

The question that remains is why the observed NIR polarization fraction for J04395 is so low, given the
much higher values displayed by objects projected to be nearby on the sky and with
similar apparent extinctions.

\end{document}